\documentclass[11pt,letterpaper]{article}
\usepackage{amsmath, cite,amssymb,color,feynmf}

\newcommand{\tmop}[1]{\operatorname{#1}}
\newcommand{\tmem}[1]{{\em #1\/}}

\definecolor{darkred}{rgb}{0.6,0.0,0.0}
\definecolor{darkblue}{rgb}{0.0,0.0,0.7}
\definecolor{darkgreen}{rgb}{0.0,0.5,0.0}
\newcommand{\blue}{\color{darkblue}}
\newcommand{\black}{\color{black}}
\newcommand{\red}{\color{darkred}}
\newcommand{\green}{\color{darkgreen}}

\newcommand{\beq}{\begin{equation}}
\newcommand{\eeq}{\end{equation}}
\newcommand{\beqa}{\begin{eqnarray}}
\newcommand{\eeqa}{\end{eqnarray}}  

\newcommand{\cL}{ {\mathcal{L}} }
\newcommand{\LFP}{ \cL_{\tmop{FP}}}

\newcommand{\mpl}{ M_{\tmop{Pl}} }
\newcommand{\Pho}{  {\red{\Phi}}_1 }
\newcommand{\PhN}{ {\green{\Phi}}_N }
\newcommand{\gPsi}{ {\green{\Psi}} }
\newcommand{\rLam}{ {\red{\Lambda}} }

\newcommand{\rLalp}{ {\red{L_\alpha}} }

\newcommand{\Hmo}{ {\blue H^m_{ {\black 1} }} }
\newcommand{\Hmt}{ {\blue H^m_{ {\black 2} }} }

\newcommand{\mbulk} { M_{{\tmop{bulk}}} }
\newcommand{\hc}{{\tmop{h.c.}}}

\begin{document}
\begin{fmffile}{msgdv2pics}

\begin{titlepage}
\begin{flushright}
CERN-PH-TH/2004-056\\
UCB-PTH-04/08\\
LBNL-54759
\end{flushright}

\vspace{15pt}

\begin{center}

{\huge\bf {\blue Massive Supergravity and Deconstruction}}

\vspace{15pt}

{
Thomas Gregoire$^a$\footnote{thomas.gregoire@cern.ch},
Matthew D. Schwartz$^b$\footnote{mdschwartz@lbl.gov}, and 
Yael Shadmi$^c$\footnote{yshadmi@physics.technion.ac.il}
}

\vspace{7pt}
{\small
${}^a$
{\it{Department of Physics, CERN, Theory Division \\
1211 Geneva 23, Switzerland\\}
}
\vspace{4pt}
${}^b$ 
{\it{Department of Physics, University of California, Berkeley,\\
      and\\
      Theoretical Physics Group, Lawrence Berkeley National Laboratory,\\
      Berkeley, CA 94720, USA\\}
}
\vspace{4pt}
${}^c$ 
{\it{Physics Department, 
Technion---Israel Institute of Technology,\\
Haifa 32000, Israel}
}
}
\end{center}

\begin{abstract}
We present a simple superfield Lagrangian for massive supergravity.
It comprises the minimal supergravity Lagrangian with interactions as well
as mass terms for the metric superfield and the chiral compensator. 
This is the natural generalization of the Fierz-Pauli Lagrangian
for massive gravity which comprises mass terms for the metric and its
trace. 
We show that the on-shell
bosonic and fermionic fields are degenerate and have the appropriate
spins: 2, 3/2, 3/2 and 1. 
We then study this interacting Lagrangian using 
goldstone superfields.
We find that a chiral multiplet of goldstones gets
a kinetic term through mixing, just as the scalar goldstone does in
the non-supersymmetric case. 
This produces Planck scale ($\mpl$) interactions
with matter and all the discontinuities and unitarity bounds associated
with massive gravity. 
In particular, the scale of strong coupling is $(\mpl m^4)^{1/5}$, where 
$m$ is the multiplet's mass. 
Next, we consider applications of massive supergravity to deconstruction. 
We estimate various quantum effects which generate non-local operators
in theory space. 
As an example, we show that the single massive supergravity multiplet
in a 2-site model can serve the function of an
extra dimension in 
anomaly mediation.

\end{abstract}

\end{titlepage}

\section{Introduction}

Only a handful of papers have been written on massive 
supergravity
\cite{Buchbinder:2002gh,Zinoviev:2002xn,Berkovits:1997zd,Berkovits:1998ua,Nojiri:2004jm}.
In fact, it seems that little is known beyond the free theory, 
and for good reason.
{\it{Massless}} supergravity is complicated enough
and at least it is constrained by powerful gauge symmetries which
a mass must break.
Moreover, the physical graviton is massless
and the gravitino cannot be, so a theory in which the graviton and gravitino
have degenerate nonzero mass cannot describe the real world. Thus, an
interacting theory of massive supergravity 
may seem to be a 
difficult and phenomenologically irrelevant mathematical exercise. 
At least, that is
the situation if one ignores recent developments in extra dimensions,
supersymmetry breaking, massive gravity and deconstruction.

First, if there are compact extra dimensions and supersymmetry, 
then there will be massive supergravity multiplets at the compactification scale.
Since it is conceivable that this scale will be accessible at future
colliders, it would be nice to have a theory for the particles which may
show up. 
As a first step in this direction, we write down
an interacting supergravity theory containing massive and massless supergravity multiplets.

But even if the extra dimensions are very small,
massive supergravity may play a crucial role in the low-energy theory.
This is the case, for example, in
models of anomaly mediation, in which
supersymmetry breaking occurs on a hidden
brane sequestered from the standard model by some distance in a fifth
dimension
\cite{Randall:1998uk}.
The basic idea in these models
is that dangerous couplings of the
visible and hidden sectors are constrained by locality
in the fifth dimension. 
It is natural to assume that such couplings are absent at tree level,
and then loop corrections are appropriately suppressed by the size of the extra dimension.
Of course, these quantum effects must be the same in the 4D effective theory
obtained after integrating out the extra dimension.
Therefore, in the 4D theory, 5D locality must manifest itself
somehow in the regulation of divergent graviton exchange diagrams by massive
KK states.
To understand this, we need an interacting theory of massive supergravity.
And then we can apply the methods
of dimensional deconstruction
\cite{Arkani-Hamed:2001ca, Hill:2000mu, Arkani-Hamed:2001nc} which have been
developed for gauge theory. This is the subject of Sections \ref{Sdd}
and \ref{Sam}.

In addition to these phenomenologically inspired motivations, there are a
number of more theoretical questions we may ask about massive supergravity.
Massive (non-supersymmetric) gravity is a very peculiar theory, and we might
hope that by adding supersymmetry we might gain insight into, or resolve some
of these peculiarities. For example, there is a factor of 4/3 difference
between tree level processes involving a massless graviton as compared to a
graviton with an infinitesimally small mass
\cite{vanDam:1970vg, zakharov}.  
So we may ask: if light passed by
a 
supersymmetric sun,
would its angle of deflection change discontinuously as
we take the mass of the supergravity multiplet to zero? In general, we would
like to know whether each such classical phenomenon has a supersymmetric
analog. 

Massive gravity is also fascinating, theoretically, at the quantum level
\cite{Arkani-Hamed:2002sp}. 
For example, if we try to discretize a gravitational dimension, consistency
requirements in the quantum theory force us to keep the lattice spacing much
larger than the Planck length
\cite{Arkani-Hamed:2003vb,Schwartz:2003vj}. To approach the continuum limit, we must
explicitly add interactions among distant lattice sites. One might hope that if
supersymmetry is essential for regulating gravity, it would manifest its
influence on some of these classical or quantum phenomena. In Section \ref{Sints} we
consider this possibility, but show that adding supersymmetry does not make
massive gravity any less eccentric.

Now, to answer any physical question about massive supergravity, we need an
interacting Lagrangian. As a start, we should look for a free theory
guaranteeing the correct on-shell degrees of freedom
\cite{Buchbinder:2002gh,Zinoviev:2002xn,Berkovits:1997zd,Berkovits:1998ua}. 
We prefer to use superfields, as in 
\cite{Buchbinder:2002gh}, rather than
components because we expect the mass term to preserve global $N=1$
supersymmetry. 
However, we will not use the Lagrangians of \cite{Buchbinder:2002gh}. 
These involve auxiliary fields not present in
any known off-shell formulation of massless supergravity, and therefore they
are difficult to generalize to the interacting case. In particular, it is hard
to tell which symmetries are broken by the mass term because we do not know
how all the fields transform under the symmetry group of supergravity. This
issue will be explored in Section \ref{Snonsusy}, and we reproduce and examine the work
of \cite{Buchbinder:2002gh} in Section \ref{Saltfree}.

We construct an improved linearized Lagrangian in Section \ref{Slinear} using the
simple and powerful formalism of superspin projectors
\cite{Sokatchev:1975gg, Siegel:1981ec}.
Conveniently, it ends
up being exactly the linearized Lagrangian for old minimal supergravity with
the addition of mass terms for the metric superfield ${\blue H_m}$ and the
chiral compensator $e^{{\blue \Sigma}}$. The compensator's mass requires the
introduction of a real superfield ${\blue P}$, which acts as a prepotential: 
${\blue \Sigma} = - \frac{1}{4} \bar{D}^2 {\blue P}$. And the action, once we
add the interactions of massless supergravity, can be written as:
\begin{equation}
  S = \int d^8 z \left\{ \mathcal{E}^{-1}[ {\blue H_m},{\blue \Sigma}] + m^2 {\blue H_m^{{\black 2}}} +
  \frac{9}{4} m^2 {\blue P}^2 \right\} \quad . \label{smsg}
\end{equation}
Here, $\mathcal{E}^{-1}[{\blue H_m},{\blue \Sigma}]$ is inverse 
of the supervierbein superdeterminant, written as
a functional of ${\blue H_m}$ and ${\blue \Sigma}$. 
For $m = 0$, \eqref{smsg} reduces to the 
action for massless supergravity \cite{Buchbinder:qv,Gates:nr}.

Already, one can see the relevance of massive supergravity to 5D physics. The
action \eqref{smsg} shares many features with a recent parameterization of
of 5D supergravity in terms of 4D superfields \cite{Linch:2002wg}.
Indeed, this
clearly written and practical paper inspired many aspects of our current
approach. However, not only is the presentation in
\cite{Linch:2002wg} confined to the free theory, but it 
includes additional auxiliary fields whose function in the
4D effective action is unknown. Refs. 
\cite{Berkovits:1997zd,Berkovits:1998ua}, 
which derive a Lagrangian from
string theory, suffer from the same drawbacks. We found that in order to
isolate the relevant degrees of freedom for {\it just} the massive
supergravity multiplet, we had to rederive the free Lagrangian from a 4D point
of view. We will comment more on the additional fields in the section on deconstruction (\S\ref{Sdd}).

While the concise form of the action \eqref{smsg} has a certain utility, it
completely obscures the propagating degrees of freedom. Nevertheless, after
performing judicious field redefinitions
and integrating out the auxiliary fields, the action reduces to:
\begin{equation}
  S = \int d^4 x \sqrt{{\blue g}} R [ {\blue g} ] - \frac{1}{4} m^2 ( {\blue
  h_{m n}^{{\black 2}}} - {\blue h}^2 )\nonumber
\end{equation}
\begin{equation}
  - \frac{1}{4} {\blue F}_{ {\blue m n} }^2 - \frac{1}{2} m^2 {\blue A}_{{ \blue m}}^2 -
  \frac{1}{2} \varepsilon^{m n p q} {\blue \psi_m} \gamma_n
  \partial_p \bar{{\blue \psi}}_{{\blue q}} + \frac{1}{4} m 
  {\blue \bar{\psi_m}} [ \gamma_n, \gamma_m ] {\blue \psi_n} +
  \cdots \quad . \label{sonshell}
\end{equation}
This contains the correct on-shell kinetic terms for a massive supergravity
multiplet: a real spin-2, a real spin-1, and a complex Dirac spin-$3 / 2$
degree of freedom, all degenerate in mass. Indeed, the on-shell quadratic
Lagrangian is fixed completely by unitary and global supersymmetry, and so
that part of \eqref{sonshell} matches previous results
\cite{Zinoviev:2002xn}. The $\cdots$ are interactions. In contrast to
the non-supersymmetric case, some of these interactions are proportional to
$m$. We explain the significance of this is Section \ref{Scomp}, where the bosonic
and fermionic component analysis of the linearized theory is worked out in
full detail.

Note that the first line of \eqref{sonshell} makes up the interacting Fierz-Pauli
Lagrangian for massive non-supersymmetric gravity. One can see that \eqref{smsg}
is the natural supersymmetric generalization of Fierz-Pauli -- the tuning
of supersymmetric mass terms ${\blue H_m^{{\black 2}}}$ and ${\blue P}^2$ to
have the ratio $9 / 4$ serves the same function as tuning the ${\blue h_{m
n}^{{\black 2}}}$ and  ${\blue h}^2$ parts of Fierz-Pauli to have the ratio $-
1$. These particular combinations eliminate the negative energy degrees of
freedom. This can be shown through the equations of motion of either theory,
but it is much clearer when we introduce goldstones to represent the
dangerous modes. The goldstone formalism is especially useful because it
reveals the scale at which these negative norm ghosts come in to haunt us.
That is, it shows the energy scale at which the effective theory breaks down,
and below which the tuning of the mass terms in Fierz-Pauli, and in (1) are
technically natural. In both cases it is at energy 
$\Lambda_5 = (\mpl m^4 )^{1/ 5}$.

The goldstones are the degrees of freedom that the massive theory has
and the massless theory lacks. Essentially, in the massless theory, we can set
them to zero by a gauge transformation. So to understand the goldstones, we
need to comprehend and to parameterize the invariance of massless
supergravity. There are many ways to do this, none of which we find to be
particularly natural or transparent. In Section \ref{Snonlinear} we discuss some of the
relevant issues, and present the set of transformations which we have found to be
most straightforward. This lets us isolate the strongest interactions, 
which turn out to be among a chiral multiplet of goldstones.
This multiplet contains the
scalar longitudinal mode of the graviton, as well as the longitudinal modes of
${\blue A_m}$ and $\bar{ {{\blue \psi}} }_{{\blue m}}$ from \eqref{sonshell}.

In Section \ref{Sdd} we show how to use the massive supergravity Lagrangian \eqref{smsg}
to deconstruct 5D supergravity theories. We work out the natural size of the
relevant UV and IR dominated contributions to operators in the effective
theory. This lets us see how well deconstruction can reproduce the effect of
sequestering within the effective theory. Section \ref{Sam} applies these rules to
a deconstructed version of anomaly mediation.

Throughout this paper, we try to stick to the supersymmetry conventions of
Wess and Bagger \cite{Wess:cp}. Also, to be clear, the terms {\it{vector}} and
{\it{scalar}} refer to the external indices on a multiplet of field (${\blue
V_m}$ is a vector) while {\it{spin-0}} and {\it{spin-1}} refer to the
irreducible representations of the Lorentz group. Contracted indices are often
omitted if there is no ambiguity, for example $( \partial {\blue A} ) =
\partial_m {\blue A_m}$ and $( \partial \partial {\blue h} ) = {\blue h_{m n,
{\black m, n}}}$. We never use curved space notation; indices are raised
and lowered with the flat space Minkowski metric $\eta_{m n}$
and with $\sigma^m_{\alpha \dot{\alpha}}$.

\section{Non-supersymmetric examples \label{Snonsusy}}
Before we begin to justify equation \eqref{smsg}, and study its properties, we will
investigate some of the critical issues in a non-supersymmetric setting. 
To begin, recall the Fierz-Pauli Lagrangian:
\begin{equation}
\LFP = 
\frac{1}{2} {\blue h_{m n}} \Box {\blue h_{m n}} + 
{\blue  h}^{ {\black{2}}}_{\blue{m n},\black{n}} 
+ {\blue h} ( \partial \partial {\blue h} )
- \frac{1}{2} {\blue h} \Box {\blue h} 
- \frac{1}{2} m^2 ( {\blue h}^{ {\black{2}}}_{\blue{m n} }  -  {\blue h}^2 ) \quad . \label{fpfree}
\end{equation}
This is the unique quadratic Lagrangian constructed out of a single two-index
tensor which propagates a spin-2 field and is free of tachyons and ghosts. It
is conveniently related to the Einstein Lagrangian expanded around flat space
${ \blue g_{m n} }  = \eta_{m n} + {\blue h_{m n} }$:
\begin{equation}
  \mpl^2 \sqrt{{\blue g}} R - \frac{1}{4} \mpl^2 m^2 ( {\blue h_{m n}^{ {\black{2}}}} -  {\blue h}^2 ) 
= \frac{1}{2}\mpl^2 \LFP + \tmop{interactions} \quad . \label{fpint}
\end{equation}
These interactions are in some sense arbitrary, as the Lagrangian no longer
has exact general-coordinate invariance (GC). In fact, \eqref{fpint} now leads to
amplitudes which violate unitarity well below $\mpl$, at energies 
$E \sim \Lambda_5 = ( \mpl m^4 )^{1 / 5}$. 
It could be better: one can raise the strong couping scale to
$\Lambda_3 = ( \mpl m^2 )^{1 / 3}$
by adding additional terms to 
\eqref{fpint}\footnote{ 
$\Lambda_3$ is the best one can do with a single spin-2 field. If the massive graviton is the
first Kaluza-Klein mode of a 5D theory, the additional KK modes raise the
strong coupling scale to $M_{5 D} = \Lambda_{2 / 3}$.}. 
However, it could also be worse:
if the tensor structure of the two derivative interactions were completely
arbitrary, that is, not protected by a custodial GC symmetry, the theory would
lead to unitarity violation at $\Lambda_7 = ( \mpl m^6 )^{1 / 7} <\Lambda_5$. 
But if the interactions in \eqref{fpint} were arbitrary, we would have a
bigger problem: when $m = 0$ the theory does not look like gravity! And so
$\mpl$ has no physical interpretation.

This last point is worth emphasizing. It is essential
that $\LFP$ looks like linearized Einstein gravity,
with its linearized symmetries, for $m = 0$.
Moreover, we also must be able to identify the auxiliary fields which appear
in the massive sector so that we understand how the symmetries are broken. For
example, consider the following Lagrangian:
\begin{equation}
\cL_A=
\frac{1}{2} {\blue h_{m n}} \Box {\blue h_{m n}} + 
{\blue  h}^{ {\black{2}}}_{\blue{m n},\black{n}} 
+ {\blue h} ( \partial \partial {\blue h} )
- \frac{1}{2} {\blue h} \Box {\blue h} \nonumber
\end{equation}
\begin{equation}
- \frac{1}{2} m^2 {\blue h_{m n}^{{\black{2}}}}
  - m {\blue A_m} ( {\blue h_{m n,{\black{n}}}} - {\blue h}_{, m} ) 
- \frac{1}{8}  ( {\blue A}_{ {\blue{m}}, n } - {\blue A}_{ {\blue{n}}, m } )^2 
+ \frac{3}{4} m^2 {\blue A}_{ {\blue{m}}}^2 \label{vecfree} \quad .
\end{equation}
At first glance, this Lagrangian seems to comprise a spin-2 and a spin-1
field. But observe that there is a bilinear coupling between them. In fact,
there are only the on-shell degrees of freedom of a massive spin-2 field; the
equations of motion force 
${\blue A_m} = {\blue h}_{ {\blue{m n}}, {\black{n}} } = {\blue h}= 0$. 
So this Lagrangian is a perfectly viable alternative to $\LFP$ \eqref{fpfree}, at the
quadratic level. However, when we embed the first line in 
$\sqrt{{\blue g}} R$
and set $m=0$, we are left with an additional free vector field with no obvious
physical interpretation. Thus, when we turn the mass back on, we are stuck in unitary
gauge. That is, we cannot study
the interacting theory using goldstones, as we do not know what symmetry
these goldstones are supposed to realize. Moreover, it is not possible to turn
\eqref{vecfree} into $L_{\tmop{FP}}$ through some straightforward field
redefinition.\footnote{
It is possible to show that \eqref{fpfree} and \eqref{vecfree} 
are dual, in
the sense that they are gauge fixed versions of a new mother Lagrangian with
an additional U(1) symmetry.}

Now, in the supersymmetric case, it is somewhat easier to produce an analog of $\cL_A$ \eqref{vecfree}
than of $\LFP$ \eqref{fpfree}. The difficulty arises because the metric is in the 
$\theta \sigma^n \bar \theta { \blue {h_{m n}}}$ component of a real vector superfield 
${\blue {H_m}}$.
So it is easy to write down the mass term
$\int d^4 \theta {\blue H}_{ {\blue m}}^{{\black 2}} = \int d^4 x {\blue h}^{{\black 2}}_{{\blue m n}}+\cdots$
but difficult to project out the trace. Thus, we can get all the quadratic ${\blue h}$ terms in $L_A$
without much work, but not those of $\LFP$. The solution, of course, is to use the conformal compensator.
In the non-supersymmetric case, it works as follows.

Start with $\LFP$ embedded in GR \eqref{fpint}. We can rewrite it slightly by
 introducing an auxiliary scalar field {\red $s$}:
\begin{equation}
  \cL_{{\red s}} = 2 \sqrt{{\blue g}} R[ {\blue{g}} ] 
- \frac{1}{2} m^2 {{\blue h}}_{{\blue{m n}}}^2+   2 m^2 {\red s}
  {\blue h} - 2 m^2 {\red s}^2 \quad .
\end{equation}
Setting ${\red{s}}$ to its equation of motion reproduces \eqref{fpint}. 
Then observe that
\begin{equation}
  \cL_{{\red s}} = 2 \sqrt{{\blue g}} R[ {\blue{g}}] - \frac{1}{2} m^2 ( e^{- 2 {\red s}} {\blue
  g_{m n}} - \eta_{m n} )^2 + 6 m^2 {\red s}^2 + \cdots
\end{equation}
\begin{equation}
  = 2 \sqrt{e^{2 {\red s}} \hat{ {\blue{g}}} } R[ e^{2 {\red s}}  \hat{{\blue g}} ] 
- \frac{1}{2} m^2 {\hat{\blue h}}_{{\blue{m n}}}^2 
+ 6  m^2 {\red s}^2 + \cdots \label{fpcc} \quad ,
\end{equation}
where ${\hat{\blue g}}_{{\blue {m n}}} = e^{- 2 {\red s}} {\blue g_{m n}}$,
${ \hat{h}}_{{ \blue {m n}}} = {\hat{\blue g}}_{{\blue {m n}}}-\eta_{m n}$,
and $\cdots$ are cubic and higher order terms. 
The awkward trace term $m^2 {\blue{h}}^2$ has been replaced by a mass
term for the conformal compensator. Moreover, the Lagrangian is now Weyl invariant when we set
$m=0$. Of course, this Weyl invariance is a complete fake, as we can just use it to set ${\red{s}}=0$. But
clearly the form \eqref{fpcc} suggests that we should be able to find a massive supergravity
Lagrangian with the features of \eqref{smsg}: mass terms for the metric and conformal compensator,
and conformal invariance in the massless limit.

\section{Constructing the Linearized Theory \label{Slinear}}
Without further introduction, we will now study the massive supergravity
Lagrangian. We will start by trying to derive a linearized Lagrangian which is
the supersymmetric version of $\LFP$. This is essentially the approach
of 
\cite{Buchbinder:2002gh}.
But since we hope to embed our linear Lagrangian in an interacting
theory, we have the additional requirement that our Lagrangian match some
known formulation of supergravity when $m = 0$.

The most (mathematically) transparent way to construct Lagrangians for
higher-spin fields is through projectors
\cite{Sokatchev:1975gg, Siegel:1981ec, Buchbinder:qv,Gates:nr}.
(See also \cite{Gates:2003cz} for a succinct application in the massless case).
For example, the spin-0 and spin-1
projectors for a Lorentz vector are:
\begin{equation}
  \omega_{m n} = \frac{\partial_m \partial_n}{\Box} \quad \tmop{and}
  \quad \theta_{m n} = \eta_{m n} - \omega_{m n} \label{vecproj}  \quad .
\end{equation}
This assures us that a Lagrangian like
\begin{equation}
  \cL = \frac{1}{2} {\blue A_m} \Box \theta_{m n} {\blue A_n} = -
  \frac{1}{4} {\blue F_{m n}^{{\black 2}}} \label{Akin}
\end{equation}
contains only spin-1 degrees of freedom. Moreover, because the projectors are
orthogonal ($\theta \omega = \omega \theta = 0 )$, this Lagrangian
automatically has a gauge invariance under 
${\blue A} \rightarrow {\blue A} + \omega \delta {\blue A}$. 
So the spin content and symmetries can basically
be read off the projectors appearing in the Lagrangian.

In supersymmetry, projectors are almost always used, whether or not they are
acknowledged. For a general scalar superfield $\gPsi$, we can isolate the
linear ($\bar{D}^2 {\green \Psi_L} = D^2 {\green \Psi_L} = 0 )$, chiral 
($\bar{D} {\green \Psi_+} = 0 )$ and antichiral ($D {\green \Psi_-} = 0 )$ sectors with:
\begin{equation}
  P_L = - \frac{1}{8 \Box} D^{\alpha} \bar{D}^2 D_{\alpha} \label{ssfproj1}
\end{equation}
\begin{equation}
  P_+ = \frac{1}{16} \frac{\bar{D}^2 D^2}{\Box} \quad \quad P_- = \frac{1}{16}
  \frac{D^2 \bar{D}^2}{\Box}  \label{ssfproj2}
\end{equation}
\begin{equation}
  P_C \equiv P_+ + P_- \label{ssfproj3} \quad .
\end{equation}
$P_C$ projects out the real chiral part ($= {\green \Phi} + \bar{ {\green \Phi}}$ for chiral
${\green \Phi}$) of a real superfield and will be very handy in what follows. For a
gauge field in a real scalar superfield ${\green V_R}$, the physical field strength
involves only $P_L {\green V_R}$ and the gauge degrees of freedom are purely $P_C {\green V_R}$.

The graviton and gravitino belong to a supergravity or superspin-$\frac{3}{2}$ multiplet.
The smallest field containing such a supergravity multiplet is a real vector superfield ${\blue H_m}$. 
Naturally, this is the
metric superfield which appears in superfield formulations of supergravity. It
has superspin components
\begin{equation}
  {\blue H_m} = \frac{3}{2} \oplus 1 \oplus \frac{1}{2} \oplus
  \frac{1}{2} \oplus 0 \quad .
\end{equation}
We can isolate these components with a set of non-local projection operators:
\begin{equation}
  \Pi^0_{m n} \equiv \omega_{m n} P_C
\end{equation}
\begin{equation}
  \Pi_{m n}^{1 / 2} \equiv \omega_{m n} P_L
\end{equation}
\begin{equation}
  \Pi^{3 / 2}_{m n} \equiv - \frac{1}{48} \sigma^{\alpha \dot{\alpha}}_m [
  D_{\alpha}, \bar{D}_{\alpha} ] [ D_{\beta}, \bar{D}_{\dot{\beta}} ]
  \sigma^{\beta \dot{\beta}}_n - \frac{1}{8} D^{\alpha} \bar{D}^2 D_{\alpha}
  \delta_{m n} - \omega_{m n} + \frac{2}{3} \Pi^0_{m n}
\end{equation}
\begin{equation}
  \Pi_{m n}^{1 / 2 ( T )} \equiv \frac{1}{48} \sigma^{\alpha \dot{\alpha}}_m [
  D_{\alpha}, \bar{D}_{\alpha} ] [ D_{\beta}, \bar{D}_{\dot{\beta}} ]
  \sigma^{\beta \dot{\beta}}_n - \Pi^0_{m n}
\end{equation}
\begin{equation}
  \Pi^1_{m n} \equiv \delta_{m n} - \Pi^0_{m n} - \Pi_{m n}^{1 / 2 ( L
  )} - \Pi^{1 / 2 ( T )}_{m n} - \Pi^{3 / 2}_{m n} \quad .
\end{equation}
These are defined for conciseness in terms of the vector projector 
$\omega_{mn}$ \eqref{vecproj} and the scalar superfield projectors \eqref{ssfproj1}-\eqref{ssfproj3}. We will not make use
of $\Pi^1_{m n}$ or $\Pi_{m n}^{1 / 2 ( T )}$ and include them
only for completeness.

\subsection{Massive Supergravity \label{Sms}}

Now we would like to choose our Lagrangian, following \eqref{Akin} as simply
\begin{equation}
  \cL = {\blue H_m} \Box \Pi^{3 / 2}_{m n} {\blue H_n} + m^2 {\blue H}_{ {\blue m}}^2 \quad .
\end{equation}
But this Lagrangian, in contrast to \eqref{Akin}, is non-local. To make it local, we
must include another projector. The simplest choice is $\Pi^0$ which leads to:
\begin{equation}
  \cL_0 = - {\blue H_m} \Box \Pi^{3 / 2}_{m n} {\blue H_n} 
+ \frac{2}{3} {\blue H_m} \Box \Pi^0_{m n} {\blue H_n} + m^2 {\blue H}_{ {\blue m}}^2 \label{lproj}
\end{equation}
\begin{equation}
  {\normalsize = \frac{1}{48} {\blue H_m} \sigma^{\alpha \dot{\alpha}}_m [
  D_{\alpha}, \bar{D}_{\alpha} ] [ D_{\beta}, \bar{D}_{\dot{\beta}} ]
  \sigma^{\beta \dot{\beta}}_n {\blue H_n} + \frac{1}{8} {\blue H_m}
  D^{\alpha} \bar{D}^2 D_{\alpha} {\blue H_m} \text{} - ( \partial {\blue
  H} )^2 + m^2 {\blue H_m^{{\black 2}}}} \quad .
\end{equation}
While this Lagrangian is local, \eqref{lproj} shows that an additional superspin-0
degree of freedom $( \Pi^0 {\blue H} )_m$ propagates, but with the opposite sign
kinetic term from $( \Pi^{3/2} {\blue H} )_m$. So one must be a ghost. The
resolution is to introduce a new auxiliary chiral field  ${\blue \Phi}$ whose
equations of motion force ${\blue \Phi} = ( \Pi^0 {\blue H} )_m = 0$. This
will be the chiral compensator. If we rewrote the Fierz-Pauli Lagrangian in
projector language, we would see that the metric trace ${\blue h}$ serves an analogous
purpose: its equations of motion force ${\blue h} = {\blue h}_{ {\blue{m n}}, m n} = 0$
\cite{Chang}.

We can isolate the ghost $( \Pi^0 { \blue H} )_m$ by defining:
\begin{equation}
  {\blue H} \equiv \partial_m {\blue H_m} = {\blue H_C} + {\blue H_L} \quad ,
\end{equation}
where ${\blue H_C^{{\black }}} = P_C {\blue H}$ is a real chiral field,
containing superspin-0, and ${\blue H_L} = P_L {\blue H}$ is linear and
contains superspin-1. Then the sector of $\cL_0$ involving ${\blue H_C}$ is:
\begin{equation}
  \cL_C = - \frac{2}{3} {\blue H_C^{{\black 2}}} - {\blue H_C^{{\black }}}
  \frac{m^2}{\Box} {\blue H_C^{{\black }}} \label{hc2} \quad .
\end{equation}
Thus, we introduce a new real chiral field ${\blue \Psi_C}$ and expand \eqref{hc2}
to
\begin{equation}
  \cL_C = - \frac{2}{3} ( {\blue H_C} - {\blue \Psi_C} )^2 - 
( {\blue  H_C} - {\blue \Psi_C} ) \frac{m^2}{\Box} ( {\blue H_C} + {\blue \Psi_C} ) \label{hcp}
\quad .
\end{equation}
It is then straightforward to see that the equations of motion for ${\blue H_C}
+ {\blue \Psi_C}$ and ${\blue H_C} - {\blue \Psi_C}$ force ${\blue H_C} =
{\blue \Psi_C} = 0$.

To represent \eqref{hcp} as a local Lagrangian note that any real chiral field can
be written as ${\blue \Psi_C} = \frac{3}{2} i ( {\blue \Sigma} - \bar{\blue \Sigma} )$, 
with ${\blue \Sigma}$ chiral. And any chiral field can be
written as ${\blue \Sigma} = - \frac{1}{4} \bar{D}^2 {\blue P}$ for real
${\blue P}$. Thus,
\begin{equation}
  {\blue \Psi_C} = \frac{3}{2} i ( {\blue \Sigma} - {\blue \bar{\Sigma}} ) = -
  \frac{3}{8} i ( \bar{D}^2 {\blue P} - D^2 {\blue P} ) \quad .
\end{equation}
Then we have, using \eqref{ssfproj2} and integrating by parts:
\begin{equation}
  \int d^8 z {\blue \Psi_C} \frac{1}{\Box} {\blue \Psi_C} = - \frac{9}{32}
  \int d^8 z D^2 {\blue P} \frac{1}{\Box} \bar{D}^2 {\blue P} =
  \frac{9}{4} \int d^8 z {\blue P}^2 \quad .
\end{equation}
So,
\begin{equation}
  \cL_C = - \frac{2}{3} ( {\blue H_C} - \frac{3}{2} i ( {\blue \Sigma} -
  {\blue \bar{\Sigma}} ) )^2 - {\blue H_C} \frac{m^2}{\Box} {\blue
  H_C} + \frac{9}{4} m^2 {\blue P}^2 \quad .
\end{equation}
The complete quadratic Lagrangian is now local:
\begin{equation}
\cL = - {\blue H_m} \Box ( \Pi^{3 / 2} {\blue H} )_m +
  \frac{2}{3} {\blue H_m} \Box ( \Pi^0 {\blue H} )_m 
+ m^2 {{\blue H}}_{{\blue m}}^2 
+ 2 i {\blue H_C} ( {\blue \Sigma} - {\blue \bar{\Sigma}} ) - 3
  {\blue \bar{\Sigma}} {\blue \Sigma} + \frac{9}{4} m^2 {\blue P}^2 
\end{equation}
\begin{equation}
=
\frac{1}{48} {\blue H^m}
  \sigma^{\alpha \dot{\alpha}}_m [ D_{\alpha}, \bar{D}_{\dot{\alpha}} ] 
[ D_{\beta},
  \bar{D}_{\dot{\beta}} ] \sigma^{\beta \dot{\beta}}_n {\blue H_n} +
  \frac{1}{8} {\blue H_m} D^{\alpha} \bar{D}^2 D_{\alpha} {\blue H_m}
  - ( \partial {\blue H} )^2 \nonumber
\end{equation}
\begin{equation}
   + 2 i ( {\blue \Sigma} - {\blue \bar{\Sigma}} ) ( \partial
  {\blue H} ) - 3 {\blue \bar{\Sigma}} {\blue \Sigma} + m^2 {\blue
  H_m^{{\black 2}}} + \frac{9}{4} m^2 {\blue P}^2
\label{comquad} \quad .
\end{equation}
In the massless limit, this is the linearized version of the old-minimal
supergravity Lagrangian with the chiral compensator 
\cite{Buchbinder:qv}.

We can check the Lagrangian by introducing goldstones. At the linearized
level, there is a set of longitudinal gauge transformations which correspond
to
\begin{equation}
  {\blue H_m} \rightarrow {\blue H_m} + \partial_m {\red G_R}
\end{equation}
for a real superfield ${\red G_R}$. Thus, ${\blue H} = ( \partial {\blue H )}
\rightarrow {\blue H} + \Box {\red G_R}$. And, because the massless part of the
Lagrangian is invariant under this gauge transformation, we can read off the
transformation of ${\blue \Psi_C}$ from the first term in \eqref{hcp}:
${\blue \Psi_C}
\rightarrow {\blue \Psi_C} + \Box {\red G_C}$. Then 
${\red G_R} = {\red G_C} + {\red G_L}$ 
comes in as
\begin{equation}
  \cL_C \supset - ( {\blue H} - {\blue \Psi_C} ) \frac{m^2}{\Box} ( {\blue
  H} + {\blue \Psi_C} )
\end{equation}
\begin{equation}
  \rightarrow - m^2 {\red G_L} \Box {\red G_L} - 2 m^2 ( {\blue H_C} -
  {\blue \Psi_C} ) {\red G_C} + \cdots \quad .
\end{equation}
This is exactly what we should expect. The linear part of ${\red G_R}$, ${\red
G_L}$, is a real multiplet with gauge invariance, so it contains the vector
longitudinal modes. These get the correct kinetic term:
\begin{equation}
  {\red G_L} \Box {\red G_L} = {\red G_L} D^{\alpha} \bar{D}^2 D_{\alpha}
  {\red G_L} \quad .
\end{equation}
The real chiral part of ${\red G_R}$, ${\red G_C}$, contains the scalar
longitudinal mode of the graviton. It gets a kinetic term from mixing with
${\blue H_C} - {\blue \Psi_C}$. So the whole goldstone formalism for
gravity, developed in
\cite{Arkani-Hamed:2002sp}
goes over beautifully to the supersymmetric case.

We will reproduce this argument more carefully in Section \ref{Snonlinear} after we have
proved the correctness of the linearized theory at the component level, and
reviewed the non-linear transformations of supergravity. But first, we will comment on some 
alternatives to \eqref{comquad}.

\subsection{Alternative Free Lagrangians \label{Saltfree}}

In a recent paper
\cite{Buchbinder:2002gh}, two candidate (free) Lagrangians for massive gravity were
suggested, neither of which match \eqref{comquad}.
Since these Lagrangians were only presented at the linearized level,  it
is unclear how to add interactions in a consistent way. In this section, we
rederive them and explain why they are not promising candidates for a non-linear theory.

We observed above that the simple Lagrangian involving only the superspin-$\frac{3}{2}$
projector is non-local, so we must include other superspin components, for
example $\Pi^0$ or $\Pi^1$. Let us allow for an arbitrary linear combination of
these two projectors. Then the general local Lagrangian is:
\begin{equation}
  \cL_H = - {\blue H_m} \Box \Pi^{3 / 2}_{m n} {\blue H_n} + \alpha \frac{2}{3}  {\blue H_m} \Box \Pi^0_{m n} {\blue H_n} 
+ ( 1 - \alpha ) \frac{2}{3} {\blue  H_m} \Box \Pi^{1 / 2}_{m n} {\blue H_n} 
+ m^2 {\blue H_m^{{\black 2}}}
\end{equation}
\begin{equation}
  = L_{3 / 2} + ( 1 - \alpha ) \frac{2}{3} {\blue H_L} {\blue H_L} - \alpha
  \frac{2}{3} {\blue H_C} {\blue H_C} - {\blue H} \frac{m^2}{\Box} {\blue
  H} \quad ,
\end{equation}
where ${\blue H} = \partial_m {\blue H_m}= {\blue H_L}+{\blue H_C}$ are
the same objects from Section \ref{Sms}.
Now we want to introduce a real auxiliary superfield ${\blue\Psi_R}$ 
so that the equations of motion enforce ${\blue H} = {\blue \Psi_R} = 0$.
Then the most general local Lagrangian involving ${\blue H}$ and ${\blue \Psi_R}$ is:
\begin{equation}
  \cL_{H \Psi}  = \frac{2 ( 1 - \alpha )}{3} {\blue H} P_L {\blue H} -
  \frac{2 \alpha}{3} {\blue H} P_C {\blue H} - \frac{m^2}{\Box} {\blue
  H}^2 + \frac{4}{3} {\blue \Psi_R} {\blue H} \nonumber
\end{equation}
\begin{equation}
+ \frac{1}{2} \beta_1 {\blue
  \Psi_R} \Box P_L {\blue \Psi_R} + \frac{1}{2} \beta_2  {\blue \Psi_R} \Box P_C
  {\blue \Psi_R} + \frac{1}{2} \gamma {\blue \Psi_R}^2 \nonumber
\end{equation}
Substituting the equation of motion for ${\blue \Psi_R}$ into that of ${\blue H}$ gives:
\begin{equation}
  ( \beta_1 \Box P_L + \beta_2 \Box P_C + \gamma ) ( \frac{4 ( 1 - \alpha
  )}{3} \Box P_L - \frac{4 \alpha}{3} \Box P_C - 2 m^2 ) {\blue H} =
  \frac{16}{9} \Box {\blue H} \quad .
\end{equation}
Since $P_L P_C = P_C P_L = 0$ and $P_C + P_L = 1$ there are two distinct
solutions which force ${\blue H} = {\blue \Psi_R} = 0$. 
The first has $\alpha = 0$ and so only
involves the $\Pi^1$ projector:
\begin{equation}
  \cL_{H \Psi}^{\alpha = 0} = \frac{2}{3} ( {\blue H_L} + {\blue \Psi_L} )^2 -
  \frac{m^2}{\Box} {\blue H_L^{{\black 2}}} - \frac{1}{\Box} \left( m
  {\blue H_C} - \frac{2}{3 m} \Box {\blue \Psi_C} \right)^2 + \frac{2}{3}
  {\blue \Psi}_{{\blue C}}^2
\end{equation}
\begin{equation}
= \frac{1}{48} {\blue H_m} \sigma^{\alpha
  \dot{\alpha}}_m [ D_{\alpha}, \bar{D}_{\alpha} ] [ D_{\beta},
  \bar{D}_{\dot{\beta}} ] \sigma^{\beta \dot{\beta}}_n {\blue H_n} +
  \frac{1}{8} {\blue H_m} D^{\alpha} \bar{D}^2 D_{\alpha} {\blue H_m}
  - \frac{1}{3} ( \partial_m {\blue H_m} )^2 + m^2 {\blue H}_{{\blue m}}^2
\end{equation}
\begin{equation}
  + \frac{4}{3} {\blue \Psi_R} \partial_m {\blue H^m} 
+ \frac{2}{3} {\blue  \Psi}_{{\blue R}}^2 - {\blue \Psi_R} \frac{\{ D^2, \bar{D}^2 \}}{36 m^2} {\blue \Psi_R} \quad .
\end{equation}

The second has $\alpha = 1$ and only involves the $\Pi^0$ projector:
\begin{equation}
  \cL_{H \Psi}^{\alpha = 1} = - \frac{2}{3} ( {\blue H_C} -
  {\blue \Psi_C} )^2 - \frac{m^2}{\Box} {\blue H_C^{{\black 2}}} -
  \frac{1}{\Box} \left( m {\blue H_L} - \frac{2}{3 m} \Box {\blue
  \Psi_L} \right)^2 - \frac{2}{3} {\blue \Psi_L^{{\black 2}}}
\end{equation}
\begin{equation}
= \frac{1}{48} {\blue H_m} \sigma^{\alpha
  \dot{\alpha}}_m [ D_{\alpha}, \bar{D}_{\alpha} ] [ D_{\beta},
  \bar{D}_{\dot{\beta}} ] \sigma^{\beta \dot{\beta}}_n {\blue H_n} +
  \frac{1}{8} {\blue H_m} D^{\alpha} \bar{D}^2 D_{\alpha} {\blue H_m} - (
  \partial_m {\blue H_m} )^2 + m^2 {\blue H}_{{\blue m}}^2
\end{equation}
\begin{equation}
  + \frac{4}{3} {\blue \Psi_R} \partial_m {\blue H^m} 
- \frac{2}{3} {\blue  \Psi}_{{\blue R}}^2 - {\blue \Psi_R} \frac{D^{\alpha} \bar{D}^2 D_{\alpha}}{18 m^2} {\blue
  \Psi_R} \quad .
\end{equation}
These are the two solutions given in 
\cite{Buchbinder:2002gh}.

Let us consider these two solutions in turn. For the first solution, $\alpha =0$, 
if we take $m \rightarrow 0$ then ${\blue \Psi_C}$ decouples. We are left with
a linearized supergravity theory with a vector auxiliary field. As discussed
in \cite{Buchbinder:2002gh} 
and
\cite{Gates:2003cz}
there is no known interacting theory of supergravity with this
auxiliary multiplet structure. It is similar to new-minimal supergravity, and
equivalent at the linearized level. But we do not know
how to add interactions in a consistent way and we cannot approach the
non-linear theory without solving the much harder problem of generating a new
formulation of supergravity.

For the $\alpha = 1$ theory, the $m \rightarrow 0$ limit decouples ${\blue
\Psi_L}$ and we are left with old-minimal supergravity with the chiral
compensator set to zero. So in this case, we can in principle consider an
interacting theory. However, the component form has a bosonic sector which is
essentially equivalent to
\eqref{vecfree}. There is a propagating vector field which
simply does not decouple. In superfield terms, we have a real field ${\blue \Psi_R}$
which is not related in any way to the auxiliary fields of the massless
sector.

\section{Components \label{Scomp}}
To show that the formal procedure of the previous section is mathematically
sound, and to get a more practical understanding of the Lagrangian, we will
now work out the components of equation (\ref{comquad}). We use the notation $|$ to indicate evaluation at 
$\theta=\bar{\theta}=0$.

For the metric superfield we define
\begin{equation}
  {\blue H^n} | = {\blue A^n} \quad 
- \frac{1}{4} D^2 {\blue H^n} | = {\blue  F^n}  \quad 
- \frac{1}{4}\sigma^{\alpha\dot{\alpha}}_m [ D_\alpha, \bar{D}_{\dot \alpha} ] {\blue H^n} | = {\blue  V^n_m} \quad 
\frac{1}{32} \{ D^2, \bar{D}^2 \} {\blue H^n} | = {\blue D^n} +  \frac{1}{2} \Box {\blue A^n} \nonumber
\end{equation}
\begin{equation}
  \frac{i}{16} \bar{\sigma}^{m \dot{\beta} \beta} \bar{D}^2 D_{\beta} 
{\blue H_{\alpha \dot{\beta}}} | = {\blue \psi_{\alpha}^m} \quad
  D_{\alpha} {\blue H_m} | =  {\blue \chi_{\alpha m}}
\end{equation}
and pull off the symmetric, antisymmetric and trace parts of ${\blue V_m^n}$ 
\begin{equation}
  {\blue V_{n m}} = {\blue v_{m n}} + {\blue \omega_{m n}} + \frac{1}{4}
  \eta_{m n} {\blue h}
\end{equation}
\begin{equation}
{\blue v_{m n}} - {\blue v_{n  m}}
= {\blue \omega_{m n}} + {\blue \omega_{n m}}
= {\blue v_{m m}} = 0 \quad .
\end{equation}
For the prepotential of the chiral compensator, we define
\begin{equation}
  {\blue P} | = {\red p} \quad - \frac{1}{4} \bar{D}^2 {\blue P} | =
  \frac{1}{2} ( {\red s} + i {\red t} ) \quad - \frac{1}{4} \sigma_m^{\alpha \dot{\alpha}}[ D_\alpha, \bar{D}_{\dot{\alpha}} ]
  {\blue P} | = {\red b_m} \quad \frac{1}{32} \{ D^2, \bar{D}^2 \} {\blue P} |
  = {\red d}
\end{equation}
\begin{equation}
  D_{\alpha} {\blue P} | = i {\red \lambda_\alpha} \quad 
- \frac{1}{4} D_{\alpha}  \bar{D}^2 {\blue P} | = {\red \zeta_\alpha} \quad .
\end{equation}
Then, after an invigorating calculation, the linearized Lagrangian \eqref{comquad} turns into:
\begin{equation}
  \cL = \frac{1}{2} {\blue v_{m n}} \Box {\blue v_{m n}} + (\partial_m {\blue
  v_{m n}})^2 - \frac{1}{6} ( {\blue h} - 6 {\red s} ) ( \partial
  \partial {\blue v} ) - \frac{1}{48} ( {\blue h} - 6 {\red s} ) \Box ( {\blue
  h} - 6 {\red s} ) \label{group1}
\end{equation}
\begin{equation}
  + \frac{4}{3} ( {\blue D_m} + \frac{1}{4} {\blue \Omega_m} - \frac{3}{4}
  {\red t}_{, m} + \frac{1}{2} \partial {\blue A}_{, m} )^2 - \frac{4}{3} | (
  \partial {\blue F} ) + \frac{3}{4} ( \partial  {\red b} ) 
 - \frac{3}{2} i {\red d} |^2 \label{group2}
\end{equation}
\begin{equation}
  + m^2 \left[ - \frac{1}{2} {\blue v_{m n}^{{\black 2}}} - \frac{1}{2} {\blue
  \omega_{m n}^{{\black 2}}} - \frac{1}{8} {\blue h}^2 + \frac{9}{8} {\red s}^2 +
  \frac{9}{8} {\red t}^2 + \frac{1}{2} {\blue A_m} \Box {\blue A_m} + 2 {\blue
  D} \cdot {\blue A} + 2 {\red } | {\blue F_m} |^2 \right.
\end{equation}
\begin{equation}
\left. +  \frac{9}{2} {\red p}
  {\red d} - \frac{9}{8} {\red p} \Box \bar{\red p}
-\frac{9}{8} {\red b}_{{\red m}}^2
  \right]
\end{equation}
\begin{equation}
  - 2 i  \bar{{\blue \psi}}_{{\blue n}} ( \bar{\sigma}_p \sigma_m \bar{\sigma}_n
  ) \partial_m  {\blue \psi_p} - ( \frac{4}{3} i {\blue \psi_n}
  \sigma_m \partial_n {\blue } {\blue \bar{\psi}_m} + h.c ) + ( -
  \frac{2}{3} i \partial_n {\blue \bar{ {\blue \chi}}_n}
  \bar{\sigma}_m \sigma_p \partial_p \bar{{\blue  \psi}}_{{\blue m}} + \hc )
\end{equation}
\begin{equation}
  - \frac{2}{3} i \partial_p \partial_n {\blue \bar{ {\blue
  \chi}}_n} \bar{\sigma}_p \partial_m  {\blue \chi}_m + (
  i \partial_n \bar{ {\blue \chi}}_n \sigma_m \partial_m {\red
  \zeta} + \hc ) + i\frac{3}{2} \bar{{\red \zeta}} \sigma_m \partial_m
  {\red \zeta} + ( \frac{1}{2} i \partial_{\rho} {\blue \psi_n} \sigma_p
  \bar{\sigma}_n {\red \zeta} + \hc )
\end{equation}
\begin{equation}
  + m^2 \left[ - ( i  {\blue \chi_m} \sigma_n \bar{\sigma}_m
  {\blue \psi_n} + h.c ) + \frac{9 i}{4} {\red \lambda} \sigma^m \partial_m
  {\red \bar{\lambda}} - i  {\blue \chi_m} \sigma_n \partial_n {\blue \bar{\chi}_m}
 - \frac{9 i}{4} {\red \lambda \zeta} + \frac{9 i}{4}
  \bar{{\red \lambda}} \bar{{\red \zeta}} \right] \quad ,
\end{equation}
where  ${\blue \Omega_m} = \varepsilon_{m n p q} \partial_n {\blue \omega_{p
q}}$.

Notice the convenient grouping in \eqref{group1} and \eqref{group2}. The
linearized gauge invariance of the massless sector can practically be read off the Lagrangian. If not for the
mass term, we could use this symmetry to go to a gauge where 
${\blue \omega} 
= {\blue A} = {\blue F}  ={\blue \chi}
={\red t}= {\red s }
={\red \lambda}={\red \zeta}=0$. 
However, we have broken the gauge invariance, and the best we can do is perform
field redefinitions:
\begin{equation}
  {\blue h} \rightarrow -3{\blue h} + 6 {\red s}
\end{equation}
\begin{equation}
  {\red d} \rightarrow {\red d} + \frac{1}{3} ( i \partial {\blue F} - i
  \partial \bar{{\blue F}} )
\end{equation}
\begin{equation}
  {\red b_m} \rightarrow 2 {\red b_m} - \frac{2}{3} ( {\blue F_m} +
  \bar{{\blue F}}_{{\blue m}} )
\end{equation}
\begin{equation}
  {\blue F_m} {\blue } \rightarrow {\blue F_m} {\blue } - \frac{3}{4} i {\red
  p}_{, m}
\end{equation}
\begin{equation}
  {\red t} \rightarrow {\red t} - \frac{2}{3} \partial {\blue A}
\end{equation}
\begin{equation}
  {\blue D_m} \rightarrow {\blue D_m} - \frac{1}{4} {\blue \Omega_m} -
  \frac{3}{4} {\red t}_{, m}
\end{equation}
\begin{equation}
   {\blue \psi_p} \rightarrow \frac{i}{2 \sqrt{2}} (  {\blue
  \psi^R_p} - \frac{1}{2} \sigma_p \bar{\sigma}_q  {\blue \psi_q^R} )
  - \frac{3}{4}i \sigma_p \bar{ {\red \zeta}} - \frac{1}{2} \sigma_p \partial_q
  \bar{{\blue \chi}}_{{\blue q}} \quad .
\end{equation}
Thus, using the conventional 
${\blue h_{\tmop{mn}}} = {\blue v_{m n}} + \frac{1}{4} \eta_{m n} {\blue h}$, 
replacing 
$ {\blue \chi_q} = \frac{1}{\sqrt{2}}  {\blue \psi_q^L}$, 
and rescaling 
${\red s}, {\red b_m}, {\red \lambda}, {\red d}$ and ${\red \zeta}$ 
by a factor of $\frac{2}{3}$ each, we have:
\begin{equation}
  \cL = \frac{1}{2} {\blue h_{m n}} \Box {\blue h_{m n}} + {\blue h}_{ {\blue m n},m}^2 
+ {\blue h} ( \partial \partial {\blue h} ) - \frac{1}{2}
  {\blue h} \Box {\blue h} - \frac{4}{3} {\red d}^2 - \frac{4}{3} ( \partial
  {\red b} )^2 + \frac{4}{3} {\blue D}_{{\blue m}}^2
\end{equation}
\begin{equation}
  - \frac{1}{2} \varepsilon^{m n p q} {\blue \psi_m^R} \sigma_n \partial_p
  {\blue \bar{ {\blue \psi}}^R_q}
\end{equation}
\begin{equation}
  + m^2 \left[ - \frac{1}{2} {\blue h_{m n}^{{\black 2}}} - {\blue h}^2
  \text{} + 3 {\blue h} {\red s} {\red {\black - \frac{3}{2} {\red s}^2}} + 3
   {\red d} {\red p} - 2 {\red b}_{{\red m}}^2 
- \frac{1}{2} ( {\blue F_m} - \bar{\blue F}_{{\blue m}})^2 - \frac{1}{2} {\red b_m} ( {\blue F_m}
+  \bar{{\blue F}}_{{\blue m}} ) \right.
\end{equation}
\begin{equation}
  + \frac{1}{2} {\red t}^2 - \frac{1}{2} {\blue \omega}_{{\blue m n}}^2
+ \frac{1}{2} {\blue A_m} \Box {\blue A_m} + 2 {\blue D_m} {\blue
  A_m} - \frac{1}{2} {\blue \Omega_m} {\blue A_m} + \frac{1}{2} ( \partial
  {\blue A} )^2
\end{equation}
\begin{equation}
   + \frac{1}{4} {\blue \psi_m^L} [ \sigma_n, \bar{\sigma}_m ] {\blue
      \psi^R_n}+ \hc + i {\red \lambda} \sigma_m \partial_m \bar{{\red
  \lambda}} - i {\red \lambda \zeta} + i \bar{{\red \lambda}} \bar{{\red \zeta}} 
\end{equation}
\begin{equation}
+ \left. \frac{1}{\sqrt{2}} \left({\red \zeta} \sigma_m
   {\blue \psi_m^L}+\hc \right) - \frac{i}{4}  {\blue \psi_m^L} ( \sigma_m
  \bar{\sigma}_n \sigma_p + \sigma_p \bar{\sigma}_n \sigma_m ) \partial_n
   \bar{{\blue \psi}}_{{\blue p}}^{{\blue L}} \right] \quad .
\end{equation}

The massless sector of this Lagrangian now looks like supergravity in
Wess-Zumino gauge. There is the correct kinetic term for the spin-2 and
spin-$\frac{3}{2}$ fields and the auxiliary vector and complex scalar forming
the off-shell supergravity multiplet. In this form, it is not hard to
integrate out the auxiliary fields, which leads to the on-shell quadratic
Lagrangian
\begin{equation}
  L= \frac{1}{2} {\blue h_{m n}} \Box {\blue h_{m n}} +
  {\blue h_{m n, m}^{{\black 2}}} + {\blue h} ( \partial \partial {\blue h} )
  - \frac{1}{2} {\blue h} \Box {\blue h} - \frac{1}{2} m^2 ( {\blue h_{m
  n}^{{\black 2}}} - {\blue h}^2 ) - \frac{1}{2}
  \varepsilon^{m n p q} {\blue \psi_m^R} \sigma_n \partial_p
  \bar{\blue \psi}^{{\blue R}}_{{\blue q}} \nonumber
\end{equation}
\begin{equation}
  - \frac{3}{2} m^2 
\left[ \frac{1}{4} ( {\blue A}_{ {\blue m},n} - {\blue  A}_{{\blue n}, m})^2 
+ \frac{1}{2}m^2 {\blue A}_{{\blue m}}^2 \right] 
+ m^2 \left[ - \frac{1}{2} \varepsilon^{ m n p q } {\blue \psi_m^L} \sigma_n \partial_p
  {\bar{{\blue \psi}}}^{{\blue L}}_{{\blue q}} + \frac{1}{4} {\blue \psi_m^L}
  [ \sigma_n, \sigma_m ] {\blue \psi^R_n} \right] \label{lonshellquad}
\end{equation}
After rescaling, this is \eqref{sonshell}. 

The field redefinitions that we made, like the gauge choice discussed above,
eliminate the dependence on
${\blue \omega},{\blue A},{\blue F},{\blue \chi},
{\red t},{\red s},{\red \lambda}$ and ${\red \zeta}$
in the massless sector.
This follows, of course, from the fact that gauge transformations
{\it {are}} field redefinitions.
But it also illustrates that if we perform,
as a field redefinition, the non-linear transformation which puts the fields
in a particular gauge, it will produce interactions in the massive sector of the
Lagrangian.
As we will show in the next two sections, it is much easier to study the interactions
by introducing goldstones directly in terms of superfields. 
But it is also useful to see, at least schematically, the fields responsible for strong coupling by
introducing component goldstones into the on-shell Lagrangian \eqref{lonshellquad}.

We need vector and scalar goldstones for ${\blue h_{m n} }$, a scalar goldstone for ${\blue A_m}$,
and goldstinos for ${\blue \psi^R}$ and ${\blue \psi^L}$
\begin{equation}
  {\blue h}_{{\blue m n}, n} \rightarrow {\blue h}_{{\blue m n}, n} + {\red
  \pi}^{{\red h}}_{{\red m}, n} + {\red \phi}^{{\red h}}_{, m, n}
\end{equation}
\begin{equation}
  {\blue A_m} \rightarrow {\blue A_m} + {\red \phi}^{{\red A}}_{, m}
\end{equation}
\begin{equation}
  {\blue \psi_m^{R, L}} \rightarrow {\blue \psi_m^{R, L}} + {\red \chi}^{{\red
  R, L}}_{, m} \quad .
\end{equation}
Then, putting back the $\mpl^2$  multiplying \eqref{lonshellquad}, the kinetic terms
are schematically
\begin{equation}
L\to  \mpl^2 \left\{ {\blue h} \partial^2 {\blue h} 
+ \bar{{\blue \psi}}^{{\blue  R}} \partial {\blue \psi^R} \right\}
\end{equation}
\begin{equation} 
+ \mpl^2 m^2 \left\{ {\blue A}
  \Box {\blue A} + \bar{{\blue \psi}}^{{\blue L}} \partial
  {\blue \psi_{}^L} + {\blue h} \partial^2 {\red \phi}^{{\red h}} + \bar{{\blue
  \psi}}^{{\blue L}} \partial {\red \chi}^{{\red R}} +
  \bar{{\blue \psi}}^{{\blue R}} \partial {\red \chi}^{{\red L}}
  + {\red \pi}^{{\red h}}_{{\red }} \partial^2 {\red \pi}^{{\red h}}_{{\red }}
  \right\}
  + \mpl^2 m^4 {\red \phi}^{{\red A}} \Box {\red \phi}^{{\red A}}
\end{equation}
From this we can read off the canonical normalizations. For example,
${\blue h}_c = \mpl {\blue h}$ and then because of the kinetic mixing,
 ${\red \phi}^{{\red h}}_c = \mpl m^2 {\red\phi}^{{\red h}}$. We find that
 ${\red \phi}^{{\red h}}$, ${\red \phi}^{{\red
A}}$, and ${\red \chi}^{{\red L}}$ all get factors of $\mpl m^2$ in their
normalization. This is perfectly consistent with supersymmetry, as they fall
into a chiral multiplet. If we were to work out the higher order terms in the
on-shell Lagrangian, they would include interactions of ${\red \phi}^{{\red
h}}$, ${\red \phi}^{{\red A}}$, and ${\red \chi}^{{\red L}}$ with the
characteristic scale $\Lambda_5 \equiv ( \mpl m^4 )^{1 / 5}$. But as we have mentioned,
it is significantly easier to work with superfields, as we will see in Section \ref{Sints}.

\section{Non-linear theory \label{Snonlinear}}
To study the interacting theory, we proceed in a way similar to \cite{Arkani-Hamed:2002sp}. 
The idea is to
restore local supersymmetry to the action \eqref{smsg} 
by introducing goldstone superfields. By the goldstone equivalence
theorem, these represent the longitudinal degrees of
freedom which become strongly coupled at high energy.
To introduce the goldstones, we will perform a finite supergravity transformation on \eqref{smsg}, parameterized by
a superfield ${\red L_\alpha}$.
Because \eqref{smsg} is not invariant, the Lagrangian will then depend on ${\red L_\alpha}$.
If we now interpret ${\red L_\alpha}$ 
as a field, it contains the goldstones. The new Lagrangian, with ${\red L_\alpha}$, 
will be invariant, because we can absorb an additional transformation in a shift of 
${\red L_\alpha}$. And so the
restored symmetry keeps the original supergravity fields transverse, and the strongly coupled longitudinal modes
appear in the goldstone fields.
In order to carry out this procedure, we need the finite supergravity transformations. In this section,
we attempt to motivate and review the formalism we have found most convenient
\cite{Siegel:mj} (see also \cite{Buchbinder:qv, Gates:nr, Linch:2002wg}), 
and we work out the transformations
to second order. The interactions will be studied using the goldstones in Section \ref{Sints}.

The natural starting point for non-linear supergravity is to consider
local translations of the form
\begin{equation}
  \gPsi ( x^m, \theta, \bar{\theta} ) \rightarrow \gPsi ( x^m + \delta x^m,
  \theta + \delta \theta, \bar{\theta} + \bar{\: \delta \theta} ) \label{natural} \quad ,
\end{equation}
with $\delta x^m$ real and $\delta \theta$ a complex Weyl fermion. This is the natural
generalization of general coordinate transformations to real superspace.
For constant $\delta x^m = i \theta \sigma^m \bar{\delta \theta \:} - i
\delta \theta \sigma^m \bar{\theta}$ it is just a global supersymmetry
rotation.

However, as we will justify {\it a posteriori}, it is convenient
to go further and consider local translations of {\it complex}
superspace
\begin{equation}
  z^M = ( z^m, \theta, \bar{\theta} ) \rightarrow ( z^m + \delta z^m, \theta +
  \delta \theta, \bar{\theta} + \delta \bar{\theta} ) \label{incoords} \quad ,
\end{equation}
where now 
$\delta \theta \neq \overline{ (\delta \bar{\theta} )}$.
This group
can be represented by
\begin{equation}
  z \rightarrow z' = ( e^{{\red \Lambda}} z ) \label{inexp} \quad ,
\end{equation}
where ${\red \Lambda}$ is a complex supervector field:
\begin{equation}
  {\red \Lambda} = {\blue } {\red \Lambda^M} \partial_M = {\blue } {\red
  \Lambda^m} \partial_m + {\red M^{\alpha}} D_{\alpha} + {\red
  \bar{N}_{\dot{\alpha}}} \bar{D}^{\dot{\alpha}} \label{inlambdas} \quad .
\end{equation}
While $\delta z^M = {\red \Lambda^M}$ holds only to first order, \eqref{inlambdas} is
nevertheless a faithful representation of \eqref{incoords} (more precisely, of
{\tmem{contractible}} transformations, connected to the identity).
Keep in mind that in our conventions 
${\red \Lambda^m}\partial_m = -\frac{1}{2} {\red \Lambda^{\alpha \dot{\alpha}}} \partial_{\alpha \dot{\alpha}}$.

Fields transform as
\begin{equation}
  \gPsi ( z ) \rightarrow \gPsi' ( z ) = e^{{\red \Lambda}} \gPsi (z)  \label{ztrans} \quad .
\end{equation}
Note that $( {\green \Psi_1 \gPsi_2} )' = {\green \Psi}_{{\green 1}}' {\green \Psi}_{{\green 2}}'$ because of the identity
$(e^{{\red \Lambda}} \gPsi ) = e^{{\red \Lambda}} \gPsi e^{-{\red \Lambda}}$.
Also note that we normally deal with superfields $\gPsi(x)$ which depend only on {\it{real}} superspace.
In that case, one can still define an active transformation by $\gPsi(x) \to e^{{\red \Lambda}} \gPsi(x)$
but can no longer interpret it as a passive coordinate shift.

As it is, with {\red $\Lambda^M$} arbitrary superfields, this supergeneral coordinate transformation (SUGC)  group is too large. We will now restrict it. First, we want chiral fields to remain
chiral. Since
\begin{equation}
  {\green \Phi} \rightarrow e^{{\red \Lambda}} {\green \Phi} = {\green
  \Phi} + {\red \Lambda} {\green \Phi} + \cdots \quad ,
\end{equation}
we need
\begin{equation}
  [ \bar{D}, {\red \Lambda} ] {\green \Phi} = 0 \label{bardl} \quad .
\end{equation}
This produces constraints on ${\red M^{\alpha}}$ and ${\red \Lambda^m}$ which
are solved by
\begin{equation}
  {\red \Lambda_{\alpha \dot{\alpha}}} = - 2 i \bar{D}_{\dot{\alpha}} {\red
  L_{\alpha}} \quad {\red M_{\alpha}} = - \frac{1}{4} \bar{D}^2 {\red
  L_{\alpha}} \label{cond1} \quad ,
\end{equation}
for an unconstrained spinor valued superfield ${\red L_{\alpha}}$. At this
point ${\red N_{\alpha}}$ is still unconstrained.

The transformation $\gPsi \rightarrow e^{{\red \Lambda}} \gPsi $ is for
general superfields, so if we want antichiral fields to remain antichiral we
would need
\begin{equation}
  [ D, {\red \Lambda} ] \bar{{\green \Phi}} = 0 \label{dl} \quad .
\end{equation}
To satisfy both \eqref{bardl} and \eqref{dl} would restrict the group too much (down to
non-susy GC). So we need some other way of keeping antichiral fields
antichiral. Enter the metric. We take the metric ${\blue H}$ to be a real
supervector field
\begin{equation}
  {\blue H} = {\blue H^m} \partial_m + {\blue H^{\alpha}} D_{\alpha} + {\bar{\blue H}}_{{ \blue \dot{\alpha}}}
  \bar{D}^{\dot{\alpha}} \quad ,
\end{equation}
with ${\blue H^m} = \bar{\blue H}^{{\blue m}}$ . We make
${\blue H}$ transform as
\begin{equation}
  e^{2 i {\blue H}} \rightarrow e^{2 i {\blue H}'} = e^{\bar{{\red
  \Lambda}}} e^{2 i {\blue H}} e^{- {\red \Lambda}} \quad .
\end{equation}
Now, we can define a covariantly antichiral field by
\begin{equation}
  \bar{{\green \Phi}}^{\ddagger} \equiv e^{2 i {\blue H}}
  \bar{{\green \Phi}} \rightarrow e^{\bar{{\red \Lambda}}}
  \bar{{\green \Phi}}^{\ddagger} \quad .
\end{equation}
Since \eqref{bardl} implies $[ D, \bar{{\red \Lambda}} ] \bar{{\green
\Phi}} = 0$, the antichirality of $\bar{{\green \Phi}}^{\ddagger}$ is
preserved.

Perturbatively,
\begin{equation}
  {\blue H} \rightarrow {\blue H} - \frac{i}{2} ( \bar{{\red \Lambda}} -
  {\red \Lambda} ) + [ {\red \Lambda} + \bar{{\red \Lambda}}, {\blue H} ]
  - \frac{i}{2} [ {\red \Lambda}, \bar{{\red \Lambda}} ] + \cdots \quad .
\end{equation}
So,
\begin{equation}
  {\blue H^\alpha} \rightarrow {\blue H^\alpha} +
  \frac{i}{2} ( {\red M^{\alpha}} - {\red N^{\alpha}} ) \quad .
\end{equation}
Since ${\red N^{\alpha}}$ remains arbitrary, we can use it to set
\begin{equation}
{\blue H^\alpha} = 0 \quad .\label{cond2}
\end{equation}
Thus, ${\red N_{\alpha}} = {\red M_{\alpha}}$ to first order, and the allowed SUGC transformations 
are determined completely in terms of ${\red L_{\alpha}}$. 

Although we will not need it here, for clarity and completeness, the gauge condition to next order is:
\begin{equation}
  {\red N^\alpha} 
= {\red M^\alpha} - {\blue H^m} ( \partial_m  {\red M^\alpha} )
 - \frac{i}{2} ( {\red \bar{\Lambda}^m} - {\red \Lambda^m} ) ( i \partial_m {\red M^{\alpha}} ) + \cdots \quad .
\end{equation}
In contrast, the other gauge constraint \eqref{bardl}, $[ \bar{D}, {\red \Lambda} ]
{\green \Phi} = 0$, implies $\bar{D} e^{{\red \Lambda}} {\green \Phi} = 0$
so \eqref{cond1} does not get further corrections.

In summary, with the conditions \eqref{bardl}  and \eqref{cond2},
the group of transformations $z \to (e^\rLam z)$ we want are determined completely by a 
spinor superfield $\rLalp$:
\begin{equation}
\rLam =  
( i \bar{D}_{\dot{\alpha}} \rLalp )\partial_{\alpha \dot{\alpha} }
+(-\frac{1}{4} \bar{D}^2 {\red L^\alpha}) D_\alpha
+(-\frac{1}{4} D^2 {\bar {\red L}}_{\red \dot {\alpha}}  +\cdots) \bar{D}^{\dot{\alpha}} \label{lamsumm} \quad .
\end{equation}
This group is a subgroup of complex SUGC representing superconformal transformations.
At the linearized level, it is also the gauge invariance associated to a
massive superspin-$\frac{3}{2}$ multiplet.
As we discussed in Section \ref{Slinear},
it is impossible to write down a local quadratic Lagrangian involving only superspin-$\frac{3}{2}$.
There, the chiral compensator came in as a Lagrange multiplier forcing the other spins to vanish
on shell. In the massless theory, one can use up some of the superconformal invariance to set
the compensator to zero. But the essential relation between superconformal transformations
and the superspin-$\frac{3}{2}$ field of supergravity remains.
From now on, we will take SUGC to mean just these symmetries, \eqref{lamsumm}.

The non-linear generalization of constructing quadratic Lagrangians with linear invariance is constructing
invariant integrals.
As in the non-supsersymmetric case, the integral of a scalar is not invariant, and
so we need to introduce scalar densities, the analog of $\sqrt{{\blue g}}$. But, in supergravity,
it is difficult to construct a density out the metric ${\blue H^m}$ and so we simply introduce
the chiral compensator for this purpose. The transformation of a supersymmetric scalar density,
to first order, is described by right action of a supervector field,
\begin{equation}
  \gPsi \overleftarrow{{\red \Lambda}} \equiv 
( - 1 )^{\varepsilon_M} \partial_M ( {\red \Lambda}^M \gPsi )
=\partial_m( {\red \Lambda^m} \gPsi) 
-D^\alpha ( {\red M}_\alpha \gPsi) 
- \bar{D}_{ \dot{\alpha}} ( \bar{{\red N}}^{{\red \dot{\alpha}}} \gPsi) \quad .
\end{equation}
This is a total derivative, and can be used to construct invariant Kahler potentials.
If we only integrate over $d^2 \theta$ the $\bar{D}$ term is not invariant and we must use
\begin{equation}
  {\red \Lambda_{\tmop{CH}}} \equiv {\red \Lambda^m} \partial_m + {\red M^{\alpha}} D_{\alpha} \quad.
\end{equation}
So, for the non-linear theory, we define the chiral compensator $e^{3 {\red {\blue \Sigma}}}$ as a
chiral density with the transformation law
\begin{equation}
  e^{3 {\red {\blue \Sigma}}} \rightarrow e^{3 {\red {\blue \Sigma}}} 
e^{ \overleftarrow{{ \red  \Lambda}}_{\red{\tmop{CH}}}} \quad .
\end{equation}
Note that for a chiral field ${\red \Lambda} {\green \Phi} = {\red
\Lambda_{\tmop{CH}}} {\green \Phi}$. This lets us construct invariant
superpotentials. Indeed, one can check that for a chiral superpotential
${\green W}$, 
\begin{equation}
( e^{3 {\red {\blue \Sigma}}} e^{ \overleftarrow{{ \red  \Lambda}}_{\red{\tmop{CH}}}} )
 ( e^{{\red \Lambda}} {\green W} ) 
= (e^{3 {\red {\blue \Sigma}}} {\green W}) 
 e^{ \overleftarrow{{ \red  \Lambda}}_{\red{\tmop{CH}}}} \quad ,
\end{equation} 
and so  $\int d^2 \theta e^{3 {\red {\blue \Sigma}}} {\green W}$ is invariant. There is also a way to construct
Kahler potentials which are invariant under finite transformations,
involving both ${\blue H}$ and $e^{3 {\red {\blue \Sigma}}}$, but it is more complicated and we will not
present it here.

Expanding to second order, the transformations are:
\begin{equation}
  \delta {\blue H_{\alpha \dot{\alpha}}} 
= \bar{D}_{\dot{\alpha}} {\red  L_{\alpha}} 
+ i \bar{D}^{\dot{\beta}} {\red L^{\beta}} \partial_{\beta  \dot{\beta}} {\blue H_{\alpha \dot{\alpha}}} 
- \frac{1}{4} (  \bar{D}^2 {\red L^{\beta}} ) D_{\beta} {\blue H_{\alpha \dot{\alpha}}}
- i {\blue H}^{\beta \dot{\beta}} \partial_{\beta \dot{\beta}} \bar{D}_{\dot{\alpha}} {\red L_{\alpha}} \nonumber
\end{equation}
\begin{equation}
  - i \sigma^{m \dot{\beta} \beta} ( \bar{D}_{\dot{\beta}} {\red L_{\beta}}
  {\red } ) \partial_m D_{\alpha} \bar{\red L}_{{\red \dot{\alpha}}} 
- \frac{1}{4}  ( \bar{D}^2 {\red L^{\beta}} ) D_{\beta} ( \bar{D}_{\dot{\alpha}} {\red
  L_{\alpha}} - D_{\alpha} \bar{\red L}_{ {\red \dot{\alpha}}} ) + \hc \label{hnl}
\end{equation}
\begin{equation}
  \delta {\red {\blue \Sigma}} = \frac{1}{12} \bar{D}^2 D_{\alpha} {\red
  L^{\alpha}} + \frac{1}{96} \bar{D}^2 ( {\red L^{\alpha}} D_{\alpha} {\red }
  \bar{D}^2 D_{\beta} {\red L^{\beta}} ) - \frac{1}{4} \bar{D}^2 ( {\red
  L^{\alpha}} D_{\alpha} {\blue \Sigma} ) \label{signl}
\end{equation} 
\begin{equation}
  \delta {\blue P} = - \frac{1}{3} D_{\alpha} {\red L^{\alpha}} - \frac{1}{24}
  ( {\red L^{\alpha}} D_{\alpha} {\red } \bar{D}^2 D_{\beta} {\red L^{\beta}}
  ) + ( {\red L^{\alpha}} D_{\alpha} {\blue \Sigma} ) + \hc \label{pnl} \quad .
\end{equation}

\section{Interacting Goldstone Superfields \label{Sints}}
To study the interacting theory, we will now introduce goldstones. Because
we have the transformation laws entirely in terms of superfields, we can study
the interactions in a manifestly supersymmetric way. All of the goldstones
fit into the spinor-valued real superfield ${\red L_{\alpha}}$. The
strong interactions involve the longitudinal modes of the fields in
this multiplet, which we can isolate by writing 
${\red L_{\alpha}} = D_{\alpha} {\red G_R}$ for real ${\red G_R}$.
And the scalar longitudinal modes are contained in the
real chiral part of ${\red G_R}$, which we can isolate with 
${\red G_R} \to {\red G_R} + {\red G} +\bar{ {\red G}}$.
This ${\red G}$, without a subscript, is a chiral field.
Therefore, we project out the strongest modes by setting:
\begin{equation}
  {\red L_{\alpha}} 
= \frac{i}{2} D_{\alpha} {\red G} \quad \bar{ {\red L}}^{ \dot{ {\red \alpha}}} 
= - \frac{i}{2} \bar{D}^{\dot{\alpha}} \bar{{\red G}},
\quad
{\bar D}{\red G} = 0 \quad .
\end{equation}
This leads to the very simple form for the linear gauge transformations \eqref{lamsumm}
\begin{equation}
  {\red \Lambda} = -2 i( \partial^m {\red G} )\partial_m \quad .
\end{equation}
These are coordinate transformations of the form $\delta x^m = \partial^m
{\red G} {\blue }$ which include the scalar longitudinal transformation,
$\delta {\blue h_{m n}} = {\red G}^{ {\red h}}_{,m,n}$ leading to strong coupling in gravity. In fact,
this chiral multiplet effectively contains the component goldstones 
$ {\red \phi}^{{\red h}}$,
$ {\red \phi}^{{\red A}}$ and 
${\red \chi}^{{\red L}}$ which we began to analyze at the end of Section \ref{Scomp}.

The second order transformations \eqref{hnl} and \eqref{pnl} are:
\begin{equation}
  \delta {\blue H^m} = \partial^m ( {\red G} + \bar{{\red G}} ) -
  2 i {\red } \partial^n {\red G} \partial_n \partial_m
  \bar{{\red G}} + 2 i {\red } \partial^n \bar{{\red
  G}} \partial_n \partial_m {\red G} - 2 i \partial^n ( {\red G} -
  \bar{{\red G}} ) \partial_n {\blue H^m} + 2 i
  {\blue H^n} \partial_n \partial_m ( {\red G} -
  \bar{{\red G}} ) \nonumber
\end{equation}
\begin{equation}
  \delta {\blue P} =  \frac{i}{6} D^2 {\red G} + \frac{i}{2} D^\alpha {\red G} D_\alpha \bar{D}^2 {\blue P}  
- \frac{1}{6} D^\alpha {\red G} \Box D_\alpha {\red G} + \hc \quad .
\end{equation}
So, at quadratic order, the mass terms become
\begin{equation}
  m^2 \int  {\blue H}_{ {\blue m}}^2 + \frac{9}{4}{\blue P}^2
\to  m^2 \int ( {\blue H_m} + \partial^m ( {\red G} + \bar{{\red G}} ) )^2 
+ \frac{9}{4} ( {\blue P} + \frac{i}{6} D^2 {\red G} 
- \frac{i}{6} \bar{D}^2 \bar{{\red G}} )^2 \quad .
\end{equation}
After an integration by parts and
an application of the identity 
$\bar{D}^2 D^2 {\red G} = 16 \Box {\red G}$ 
this simplifies to
\begin{equation}
  m^2 \int {\blue H_m^{{\black 2}}} + \frac{9}{4} {\blue P}^2 - 2 (
   \partial {\blue H} - \frac{3i}{2} ( {\blue \Sigma} - \bar{ {\blue \Sigma}} ) ) 
 ( {\red G} +  \bar{{\red G}} ) \quad .
 \end{equation}
As expected, the kinetic term for
${\red G}$ has vanished, and ${\red G}$ picks up a kinetic term
from mixing. Moreover, it mixes with the same combination of metric and compensator that
${\blue \Sigma}$ itself does in the massless Lagrangian, \eqref{comquad}. 
So we can undo the mixing with a Weyl transformation ${\red \Sigma} \to {\red \Sigma} - i {\red G}$,
producing an an ordinary $\bar{\red G} {\red G}$ kinetic term. This is exactly analogous to how the kinetic
mixing works in non-supersymmetric massive gravity.
Note that this Weyl transformation will produce a coupling of
${\red G}$ to matter with $\mpl$ strength. So this coupling will remain even in the limit $m \rightarrow 0$.
Therefore, there is a supersymmetric analog of the vDVZ discontinuity
\cite{vanDam:1970vg,zakharov}.

To canonically normalize the fields, note that the metric and the compensator
have kinetic terms in the massless action, which we have defined with a
coefficient $\mpl^2$. Schematically,
\begin{equation}
\cL = \mpl^2 ( {\blue H}\partial^2{\blue H} + \bar{{\blue \Sigma}}{\blue \Sigma})
+ \mpl^2 m^2 (\partial {\blue H} +{\blue \Sigma}) {\red G} \quad .
\end{equation}
So ${\blue H}_{{\blue m}}^c = \mpl {\blue H}_{{\blue m}}$, 
${\blue \Sigma}^c = \mpl {\blue \Sigma}$
and ${\red G}^c = \mpl m^2 {\red G}$. If we had included the linear part of
the goldstone multiplet ${\red G_R}={\red G_L} + {\red G}+ \bar{\red G}$ we would
see that it gets a kinetic term from the mass without mixing and is normalized by
${\red G}_{{\red L}}^c = \mpl m {\red G_L}$. But the chiral mode, ${\red G}$, has the smallest kinetic
term and the largest interactions.
It is not hard to see that strongest interactions come from
terms cubic in ${\red G}$. Expanding the mass terms to next order
order,we have:
\begin{equation}
  m^2 {\blue H}_{ {\blue m}}^2 + \frac{9}{4} {\blue P}^2 \rightarrow 
 \frac{-2 i}{\mpl m^4}
 \left(-3 \partial_a {\red G}_{{\red c}}^\dagger \Box {\red G_c} \partial_a {\red G_c} 
- 2 \partial_a {\red G_c}^\dagger \Box {\red G_c} \partial_a {\red G}_{{\red c}}^\dagger 
+ {\red G_c} \Box {\red G_c}^\dagger \Box {\red G_c}^\dagger\right) 
+ \hc + \cdots \nonumber
\end{equation}
These vertices lead to amplitudes which violate unitarity at the energy 
$(\mpl m^4 )^{1 / 5} \equiv \Lambda_5$.
So we take the cutoff to be $\Lambda \sim \Lambda_5$.

In a consistent effective field theory, naive dimensional analysis
\cite{Manohar:1983md, Cohen:1997rt,Luty:1997fk}
tells us that we must include
all higher dimension operators suppressed by the cutoff.
In supersymmetric theories, the easiest way to keep the dimensions straight
is to rescale $\theta$ to be dimensionless by 
$\theta \to \frac{\hat{\theta}}{\sqrt{\Lambda}}$.
Then
canonically normalized bosonic superfields, 
such as 
$\hat{\Psi}(x,\frac{\hat{\theta}}{\sqrt{\Lambda}},
\frac{\hat{\bar{\theta}}}{\sqrt{\Lambda}})$
have mass dimension 1.
In our case, we know that because the goldstone field ${\red G}$ is
strongly coupled at $\Lambda$, we must include operators in the Kahler potential like
\begin{equation}
\frac{\Lambda^2}{16 \pi^2}
\left( \frac{\partial}{\Lambda} \right)^a \left( \frac{4 \pi \hat{{\red G}}_c }{\Lambda} \right)^b
= 
\frac{\Lambda^2}{16 \pi^2}
\left(\frac{\partial}{\Lambda} \right)^a \left( \frac{4 \pi \mpl m^2 \hat{{\red G}}}{\Lambda^3} \right)^b \quad .
\end{equation}
We have included a $(16 \pi^2)^{-1}$ in front as a loop factor,
and a $4 \pi$ next to ${\red G}$ for canonical normalization. Matching the $a=6$, $b=3$ term to the tree level expression
enforces $\Lambda < (4 \pi)^{1/5} \Lambda_5 \sim 1.6 \: \Lambda_5$. Since we are neglecting factors of order 1, it is consistent
to ignore the $4 \pi$'s in the our analysis.

Now, ${\red G}$ comes out of ${\blue H}$, so the corresponding unitary gauge term is
\begin{equation}
\Lambda^2 \left(\frac{\partial}{\Lambda} \right)^a 
\left( \frac{\mpl m^2}{\Lambda^3}\frac{\hat {{\blue H}}_c}{\mpl} \right)^b
=\Lambda^2 \left(\frac{\partial}{\Lambda} \right)^a 
\left( \hat{e} \frac{\hat{\blue H}_c}{\Lambda} \right)^b \quad .
\end{equation}
The important point is that the couplings of the transverse modes do not enter as 
$\hat{\blue H}_c/\Lambda$ but are down by a weak coupling factor of
\begin{equation}
\hat{e} \equiv \frac{m^2}{\Lambda^2} = \left(\frac{m}{\mpl}\right)^{\frac{2}{5}} \quad .
\end{equation}

\section{Dimensional Deconstruction \label{Sdd}}
Now that we understand the Lagrangian for massive supergravity, we can string together an extra dimension. This works essentially the same way for supergravity as it does for gauge theories. To construct an $N$-site model, we begin with $N$ independent 4D supergravity theories:
\begin{equation}
  S = \int d^4 x d^4 \theta \sum_j M_j^2 {\mathcal E}^{- 1}[{\blue H}^{{\blue m}}_j,{\blue \Sigma}_j]  \quad .
\end{equation}
$M_j$ are the Planck scales on each site, which we may take to be distinct.
Note that there is only one set of coordinates, but the Lagrangian is
invariant under $N$ copies of SUGC.
Now we add nearest-neighbor interactions which break SUGC$^N$ down to
the diagonal SUGC subgroup:
\begin{equation}
\label{decon}
  L_U = \sum_j M_j^2 {\mathcal E}^{-1}[{\blue H}^{{\blue m}}_j,{\blue \Sigma}_j] 
+  M_j^2 m_j^2  \left\{ ({\blue H}^{{\blue m}}_{j+1} - {\blue H}^{{\blue m}}_j )^2 
+ \frac{9}{4} ( {\blue  P}_{j + 1} - {\blue P}_j )^2 + \cdots_j \right\} \quad .
\end{equation}
Here, $m_j$ characterize the masses. 
If we set $M_j = M$ and $m_j = m$ 
then the theory contains a tower of massive supergravity multiplets with
couplings $\mpl^2 = N M^2$ and masses 
$m_n = m \sin{\frac{n\pi}{N}}$,
with $n = 0 \ldots N - 1$. For large $N$ this
approximates one KK tower of a compactified 5D supergravity theory, as should
be expected for a discretization.
Also, $\cdots_j$ stands for the terms needed to make the mass terms invariant under
the SUGC group on site $j$. This is done by the standard procedure for adding
gravitational interactions (covariant derivatives and chiral compensator
terms) to a flat space Kahler potential.
The analog in ordinary gravity is to write
$\sqrt{g} h_{\mu \nu}^2$ instead of simply $h_{\mu \nu}^2$ in the mass term.
While we ascribe no meaning to the statement that these
additional terms are ``necessary,'' they are nevertheless helpful to guarantee
that the low energy theory has diagonal SUGC symmetry which gives it a chance
at being phenomenologically viable. 

How does sequestering work in theory space? Suppose we have some chiral field $\Pho$ on site 1 and another chiral field
$\PhN$ on site $N$. Since $\Pho$ and $\PhN$ are at different sites,
we can plausibly omit contact terms like $\Pho^\dagger \Pho \PhN^\dagger\PhN$ at tree level. But such terms will be generated by quantum corrections, and we must estimate their size to see how they are suppressed by the ``distance'' between sites 1 and $N$. Before doing that, it is instructive to restore the broken  $\tmop{SUGC}^N$ symmetry by promoting the Lagrangian to a non-linear sigma model with the introduction of compensating goldstone superfields. As we saw in Section \ref{Snonlinear}, a concise parameterization of the symmetry group can be made with a spinor superfield ${\red
L_{\alpha}}$, so we include $N - 1$ spinor superfields, one for each link. The non-linear sigma model is then
\begin{equation}
  L_{\sigma} = \sum_j M_j^2 {\mathcal E}^{- 1} [{\blue H}^{{\blue m}}_j,{\blue \Sigma}_j] 
\end{equation}
\begin{equation}
+m_j^2 M_j^2 \left\{ ({\blue H}^{{\blue m}}_{j+1} 
  - {\blue H}^{{\blue m}}_j + \bar{D}_{\dot{\alpha}}
  {\red L}^j_{{\red \alpha}} + \cdots_{NL} )^2 
+ \frac{9}{4} ( {\blue P}_{j+1} - {\blue P}_j - \frac{1}{3} D_{\alpha}
{\red L}_{{\red \alpha}}^j + \cdots_{NL} )^2 + \cdots_j \right\} \nonumber
\end{equation}
where the $\cdots_{NL}$ are the higher order, nonlinear, parts of SUGC transformations 
(cf. Eqns. \eqref{hnl} and \eqref{pnl}). Now the
Lagrangian has the full SUGC$^N$ symmetry under which the ${\red
L_{\alpha}^{{\black j}}}$ transform as bifundamentals. Explicitly (or rather,
implicitly),  the transformations are:
\begin{equation}
  e^{2 i {\blue H}_j} \rightarrow e^{\bar{\red \varepsilon}_j}
e^{2  i {\blue H}_j} e^{- {\red \varepsilon}_j}
\end{equation}
\begin{equation}
  e^{ {\red \Lambda}_j [ {\red L}_j^{{\red \alpha}} ] } 
\rightarrow e^{ {\red \varepsilon}_j }  e^{ {\red \Lambda}_j [{\red L}_j^{{\red \alpha}} ] } 
e^{- {\red \varepsilon}_{j + 1}}
\end{equation}
We are using the notation of Section \ref{Snonlinear}: the operators ${\red \varepsilon}_j$, ${\blue H}_j$
 and ${\red \Lambda}_j$ are all supervector fields, (e.g. ${\blue H} = {\blue H^m} \partial_m$) 
and ${\red \Lambda}_j[{\red L}_j^{{\red \alpha}} ]$ invokes the spinor superfield parameterization;
 ${\red \varepsilon}_j$ is the SUGC on site $j$. 
We emphasize that there are only one set of
coordinates $x, \theta, \bar{\theta}$ and all the transformations are active 
transformations under which the fields transform and the coordinates stay put.

The Lagrangians $L_U$ and $L_{\sigma}$ present the same physical theory, in
different gauges. Indeed, $L_{\sigma}$ reduces to $L_U$ after we use the extra
symmetry to go to the gauge ${\red L_{\alpha}^{{\black j}}} = 0$. But the
sigma model formulation is useful because we can now see why the
UV completion should respect locality in theory space: it must preserve the
$\tmop{SUGC}^N$ symmetry. This symmetry forbids tree-level contact terms
involving $\Pho$ and $\PhN$.
Of course, these terms are generated through quantum corrections,
but those are easy to estimate. 
There are a few types of corrections, 
which are  analogous to the corrections in the  
well-studied gauge theory case 
\cite{Arkani-Hamed:2001nc}.

First, there is the contribution from UV divergent gravity loops. 
Note that we can decouple any of the links by taking $M_j \to \infty$
holding $\Lambda_5^j = M_j m_j^4$ fixed. This works because the only interactions
with  $\Lambda_5$ strength are among the chiral goldstones ${ \red G}^j$,
coming from ${\red L}^j_\alpha = D_\alpha{\red G}^j+\cdots$ which do 
not mix sites. The proper procedure, according to naive dimensional analysis 
is to include all the operators coupling the links with their proper strengths.
As we saw in the previous section (see also
\cite{Arkani-Hamed:2002sp}), these have the form
\begin{equation}
 \Lambda^2 \left( \hat{e}_j
  \frac{\hat{{\blue H}}^c_{j + 1} - \hat{{\blue H}}^c_j}{\Lambda} \right)^p \left(
  \frac{\partial}{\Lambda} \right)^q \left( \frac{\hat{{\red G}}_j^c}{\Lambda}
  \right)^p \quad ,
\end{equation}
where $\hat{{\blue H}}^c$ and $\hat{{\red G}}^c$ are the canonically normalized metric and 
goldstone fields and
\begin{equation}
  \hat{e}_j \equiv \left( \frac{m_j}{M_j} \right)^{2 / 5} \quad .
\end{equation}
Once we use a self-consistent form of the links, we see that each vertex 
connecting nearest neighbors will have at least one factor of $\hat{e}$.
A standard spurion argument then shows that the UV contribution 
is at most
\begin{equation}
\frac{\Lambda^2}{\mpl^4}
{\hat{e}}^{2(N-1)} \Pho^2 \PhN^2 \label{uv} \quad .
\end{equation}

The second quantum contribution we are concerned with also comes from gravity loops, but it is
saturated in the IR and completely finite.
It is easy to see why the 1-loop contribution cannot vanish completely.
If we run down to low energies, below the mass of the first KK mode, 
$m_1 \sim m/N \equiv \frac{1}{R}$
the theory just appears to be 4D supergravity coupled to the chiral
fields $\Pho$ and $\PhN$, with a cutoff at $1/R$.
Since a 4D gravity loop diagram, such as
\begin{equation}
\parbox{40mm}{
\begin{fmfgraph*}(90,50)
\fmfleft{p1,p2}
\fmfright{q1,q2}
\fmf{plain}{p1,i1}
\fmf{plain}{p2,i2}
\fmf{plain}{o1,q1}
\fmf{plain}{o2,q2}
\fmf{photon,tension=.4}{i1,i2}
\fmf{photon,tension=.4}{o1,o2}
\fmf{photon,tension=.4}{i1,o1}
\fmf{photon,tension=.4,label=${\blue H}$}{i2,o2}
\fmfv{l=$\Pho$,l.a=120,l.d=.05w}{p1}
\fmfv{l=$\Pho$,l.a=-120,l.d=.05w}{p2}
\fmfv{l=$\PhN$,l.a=60,l.d=.05w}{q1}
\fmfv{l=$\PhN$,l.a=-60,l.d=.05w}{q2}
\end{fmfgraph*} }
\end{equation}
knows nothing about locality in theory space, dimensional analysis with
cutoff $1/R$
tells us that this operator must appear with coefficient
\begin{equation}
  \frac{1}{16 \pi^2 \mpl^4 R^2} \Pho^2 \PhN^2 \label{finite} \quad .
\end{equation}
The amazing thing about deconstruction is that locality in theory space
guarantees that these diagrams really are cutoff at $1 / R$
in the full theory \eqref{decon}.\footnote{With 
2 sites there is a residual logarithmic divergence.}
That is, the massive supergravity modes regulate the divergences.
Essentially, we know the diagrams are cut-off because the only field configurations
which contribute are non-local. In 5D the operator requires a Wilson loop going
around the whole space, and in deconstruction it requires extended field
configurations which can be understood with the Coleman-Weinberg potential
(again, see \cite{Arkani-Hamed:2001nc}).

Now, there is one more way the operator $\Pho^2 \PhN^2$ might appear. 
Since the theory space Lagrangian $L_U$ \eqref{decon} is
non-renormalizable, we must imagine that it is eventually embedded in a UV
completion. Already, the cutoff dependence of \eqref{uv} indicates that
the gravity sector is sensitive to UV effects
and it is certainly consistent within the effective theory to take this as the only contribution.
But, by analogy with extra dimensional theories, we can also imagine
that there are ``bulk'' states near or slightly above the cutoff, $\mbulk  \gtrsim \Lambda$,
which couple to both site 1 and site $N$ (or both branes
in a continuum 5D theory), and to $\Pho$ and $\PhN$.
In theory space, bulk states correspond to new fields
at each site, with standard nearest-neighbor hopping terms. 
Integrating these fields out at tree-level produces a third contribution to
non-local operators:
\begin{equation}
  \frac{1}{\mbulk^2} \left( \frac{N}{R \mbulk} \right)^{2 (N-1)} \Pho^2 \PhN^2 \label{bulk} \quad .
\end{equation}
In general, we expect $\mbulk$ to be much larger than 
the inverse lattice spacing $N/R$,
and so this contribution is negligible for large $N$. 
But for small $N$ it may be important.

Finally, let us formally take the continuum limit
of the Lagrangian \eqref{decon}.\footnote{This 
limit cannot actually be taken in a consistent quantum theory,
see \cite{Arkani-Hamed:2003vb, Schwartz:2003vj}.
Here we are just making observations about a correspondence between degrees
of freedom at the classical level.}
At the linearized level, we recover the 5D supergravity Lagrangian of 
\cite{Linch:2002wg}, without the radion field,
and with our goldstones ${\red L_\alpha}$ representing their
$\Psi_\alpha$. Of course, this is exactly what we expect, as
we can use 5D symmetries to set $\Psi_\alpha=0$ and then restore these
symmetries with goldstone fields after the dimensional reduction. 
Note that 
the $N=2$ supersymmetry of 5D is non-linearly realized by the goldstones,
as it must be because we have a consistent theory with two (Weyl) gravitini.

\section{Two Site Anomaly Mediation \label{Sam}}
As an example, to illustrate the simplicity of supergravity in theory
space, we sketch a 2-site model of anomaly mediation. Anomaly mediation is
a method of communicating supersymmetry breaking to the standard model. 
It relies on the fact that soft-masses for sfermions and gauginos
are automatically generated
in the presence of any supersymmetry-breaking sector that couples
to gravity, by virtue of the scale anomaly of the standard 
model \cite{Randall:1998uk,Giudice:1998xp}.
Moreover, because the soft masses are related to the breaking of scale invariance,
they are completely determined by the anomalous dimensions of 
the standard model fields.

The main advantage of anomaly-mediated supersymmetry breaking is 
that it solves the supersymmetric flavor problem. 
Because the scalar masses are   
insensitive to UV physics
\cite{Randall:1998uk, Giudice:1998xp,Pomarol:1999ie,Katz:1999uw}, 
the only flavor breaking spurions 
at low energy are the Yukawa matrices and flavor-changing neutral currents are 
naturally suppressed by the superGIM mechanism.
For this solution to work, the anomaly mediated contribution
must dominate other sources of scalar soft masses.
In particular,  soft masses  generated in the UV where there is 
flavor physics at work are dangerous and need to be suppressed.
Probably the simplest way of achieving this involves sequestering a hidden
sector with an extra dimension; if the standard model is confined to
one brane, and the supersymmetry breaking sector
to another, dangerous operators are forbidden by locality.
We can already see, by the considerations of the previous section, that all
propitious features of sequestered sector anomaly mediation should 
be reproduced in theory space.

The simplest model has two sites:
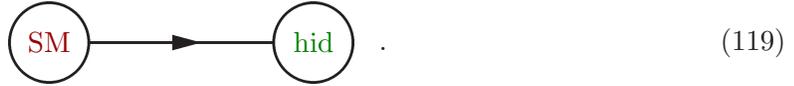
\begin{equation}
\parbox{40mm}{
\begin{fmfgraph*}(100,30)
\fmfleft{v1}
\fmfright{v2}
\fmf{fermion}{v1,v2}
\fmfv{l={\red SM},l.a=90,l.d=0.0w, 
decoration.shape=circle,decoration.filled=empty,
decoration.size=0.3w}{v1}
\fmfv{l={\green hid},l.a=270,l.d=.0w, 
decoration.shape=circle,decoration.filled=empty,
decoration.size=0.3w}{v2}
\end{fmfgraph*} } 
\quad .
\end{equation}
We put the standard model on site 1 and the hidden sector
on site 2.
Each site has its own supergravity multiplet, and
the sites only interact through a supergravity link. 
In unitary gauge, where the link
is eaten, the relevant part of the Lagrangian is
\beqa\label{amsb}
  \cL &=& \frac{1}{2} \mpl^2 {\mathcal E}^{- 1} [ {\blue H}^{{\blue m}}_1, {\blue \Sigma}_1 ] +
  \frac{1}{2} \mpl^2 {\mathcal E}^{- 1} [ {\blue H}^{{\blue m}}_2, 
{\blue \Sigma}_2 ]\nonumber \\
  &+& \mpl^2
  \frac{1}{R^2} \left\{ (\Hmt-\Hmo)^2 + \frac{9}{4} ( {\blue P}_2 - {\blue P}_1 )^2 +
  \cdots_1 \right\}\\
 &+& ( {\red Q}^\dagger {\red Q} +\ldots_1 ) + ({\green S}^\dagger {\green S} +\ldots_2)
\quad .
\nonumber
\eeqa
Here we have only shown the Kahler potential part.
${\red Q}$ stands for MSSM matter fields
and ${\green S}$ is a hidden sector field.
The $\cdots_j$ are the additional terms in the Kahler potential required to
make it invariant under the SUGC group on site $j$. This theory has a massless
supergravity multiplet 
${\blue{H^m_{{\black 1}}}}+{\blue{H^m_{{\black 2}}}}$ 
with Planck strength
$\mpl$ and a massive supergravity multiplet
${\blue{H^m_{{\black 1}}}}-{\blue{H^m_{{\black 2}}}}$ 
of mass $1 / R$. We use $R$ only to make contact with the parameters of 5D
anomaly mediation, as this 2-site model clearly has no continuum
interpretation.

We assume that the hidden sector chiral superfield ${\green S}$ gets an $F$-term 
vacuum expectation value signaling the breaking of supersymmetry.
This is communicated to the visible sector just like in 5D anomaly mediation,
through the massless supergravity mode, or equivalently the $F$ term of its
chiral compensator ${\blue \Sigma}$: $\langle F_{{\blue \Sigma}} \rangle \sim
\langle F_{\green S} \rangle / \mpl$. Just as in 5D, the gravitational
anomaly is completely independent of UV physics and so masses of the MSSM
particles are proportional to their $\beta$ functions. For example, sfermion
masses are roughly:
\begin{equation}
  m_s^2 \sim {\left(\frac{\alpha}{4 \pi}\right)}^2 
\langle F_{{\blue \Sigma}} \rangle^2  \quad .
\end{equation}
And, like in 5D anomaly mediation, some of the sleptons will be tachyonic. 
We will not attempt to solve this problem here -- 
we are merely replicating anomaly mediation in a two site model.

The next step is to consider quantum corrections
which generate contact terms between the standard model and
supersymmetry breaking sector. 
The biggest danger comes from soft masses
coming out of contact terms like
${\red Q}^\dagger {\red Q} {\green S}^\dagger {\green S}$ 
generated by UV physics.
But,
as we have
argued in the previous section, these operators are highly constrained by
locality in theory space. 
In the 2-site model, the contribution from massive fields at the cutoff 
(cf. \eqref{bulk} with $\mbulk \sim \Lambda$) 
is of order
\begin{equation}
\Delta L_{\tmop{bulk}}\sim  \frac{1}{R^2 \Lambda^4} {\red Q}^\dagger {\red Q} 
{\green S}^\dagger {\green S} \quad .
\end{equation}
And the contribution from gravity loops (cf. \eqref{uv}) is
\begin{equation}
\Delta L_{\tmop{UV}} \sim  \frac{1}{\mpl^2}
 \left(\frac{1}{\mpl R}\right)^{12/5}
{\red Q}^\dagger {\red Q} 
{\green S}^\dagger {\green S} \quad .
\end{equation}
If we take the cutoff to be around the strong coupling scale
$\Lambda \sim (\mpl R^{-4})^{1/5}$
both of these terms are suppressed by powers of $R$. 
We can easily achieve more suppression with additional sites.

In addition to UV contributions, there is also the finite IR contribution at one loop 
(cf. \eqref{finite}):
\begin{equation}
  \Delta L_{\tmop{IR}} \sim \frac{1}{16\pi^2
\mpl^4 R^2} 
 {\red Q}^\dagger {\red Q} {\green S}^\dagger {\green S} \quad .
\end{equation}
These operators can be important because they
contribute to scalar masses:
\begin{equation}
  \delta m_s^2 \propto 
\frac{1}{16 \pi^2 \mpl^4 R^2} 
\langle F_{\green S}  \rangle^2 \label{smass} \quad .
\end{equation}
The analogous continuum contribution was calculated recently 
in 
\cite{Buchbinder:2003qu}
and
\cite{Rattazzi:2003rj} 
and was found to be negative.
We should expect the same sign for \eqref{smass} and therefore
it cannot be immediately used to ameliorate the problem of tachyonic sleptons, 
as one might have hoped.

\section{Conclusions \label{Sconc}}
In the first several sections of this paper, we presented and analyzed 
a natural interacting theory of massive supergravity. It is given by the
Lagrangian \eqref{smsg}. On shell, this theory contains a massive supergravity
multiplet containing a spin-2, a spin-1 and two spin 3/2 fields all degenerate
in mass. The Lagrangian is constructed using the 4D $N=1$ superfield
formalism, so it manifestly preserves global supersymmetry. We have
demonstrated the validity of the Lagrangian in three ways: first, with a formal
analysis using spin projectors, second, using the explicit component expansion
of both the bosonic and fermionic sectors, and finally using goldstones. 
This last method is particularly useful as it leads to an
efficient way to study the interactions in the theory. Indeed, part of the
justification of our Lagrangian is that it contains only the fields of minimal
supergravity, and so the interactions and (broken) symmetries of the theory
can be simply lifted from the massless case. With other linearized
Lagrangians, such as the ones in 
\cite{Buchbinder:2002gh}, 
there are
additional auxiliary fields whose symmetry properties are unknown. Therefore,
working out the effect of the mass term on their interactions appears
prohibitively difficult. Curiously, while attempting such a program  
we came across a mysterious alternative to the Fierz-Pauli Lagrangian for a
massive graviton invoking an auxiliary vector field \eqref{vecfree}. As far as we
know, 
this Lagrangian was not presented previously,
and it may be of interest in its own right.

Returning to massive supergravity, we found by merging the goldstone formalism for
gravity developed in 
\cite{Arkani-Hamed:2002sp}
and the representation of supergeneral coordinate
transformations from \cite{Siegel:mj}
that massive supergravity has the same scale
of strong interactions as massive (non-supersymmetric) gravity. Namely, it
breaks down at $\Lambda_5 = ( \mpl m^4 )^{1 / 5}$. We worked out the form of
the strongest interactions in terms of a chiral multiplet of goldstones. This
multiplet, in unitary gauge, comprises the scalar longitudinal mode of the
graviton, the gravitino's chiral conjugate, and a scalar auxiliary field. One
might describe it as the chiral supergoldstone of a real superfield containing
the graviton's vector longitudinal polarizations, a Dirac goldstino, and a
vector auxiliary field. In Section \ref{Sints} we simply called it ${\red G}$. The
most critical feature of ${\red G}$ is that it only gets a kinetic term from
mixing with the metric superfield. The consequences of this mixing in
supergravity are just as dire as for non-supersymmetric gravity: it leads to
vDVZ type discontinuity as the mass is taken to zero; it causes the
gravitational field around massive sources to break down at distances much
larger than the Schwarzschild radius; and it prevents the reproduction of long
distance supergravitational phenomena on a lattice. 
We have not actually demonstrated any of this, but the results follow 
trivially from the kinetic mixing and the logic in \cite{Arkani-Hamed:2002sp}. 
One might have hoped that the gravitino would miraculously cancel the troublesome 
amplitudes in graviton scattering
which cause these effects, but this does not happen. One might also have hoped
to unravel the strange holographic-type bounds which turned up in the lattice
investigation of
\cite{Arkani-Hamed:2003vb,Schwartz:2003vj}, 
especially considering supersymmetry's role in
consistent theories of quantum gravity. But again, supersymmetry is of no help.

On the brighter side, the Lagrangian for massive supergravity 
naturally leads to theory space versions of sequestered sector models. 
In such models, visible and hidden sector fields are physically separated
in an extra dimension, so that couplings between them are suppressed.
We have shown that the same
dangerous operators can be 
suppressed
in a purely 4D theory by the addition of
massive supergravity modes with prescribed couplings. Locality in theory space
serves the same proscriptive function as locality in an extra dimension.
We considered in particular a 2-site model of anomaly mediation.
Working through the relevant corrections, by dimensional and symmetry
analysis, we found that deconstruction can reproduce the phenomenological success (and
failures) of a 5D anomaly mediated model. 

Our goal in the analysis of deconstruction was to exhibit the correspondence
between supergravity in 5D and in theory space, particularly in regard to
issues of locality. In this vein, deconstruction 
provides a useful tool for studying extra dimensions. But it is also true
that these theories have space to improve on the extra dimensional models,
as 4D effective theories are less restrictive. For example, the radion
can be simply thrown out. If the radion, or
some other light field, ever proves to be useful in a compactified extra
dimensional model, it can easily be co-opted into a theory space Lagrangian.
But it is equally possible that fields with no extra dimensional
interpretation at all will prove advantageous.

\section{Acknowledgements}
We would like to thank Z.~Chacko, W.~Linch, 
Y.~Nomura, R.~Rattazzi and Y.~Shirman
for helpful discussions. M.~Halpern has pointed out that 
an early work \cite{Halpern:1975yj} contains some of the elements of deconstruction.
The research of Y.S. is supported in part by the Israel Science 
Foundation (ISF) under grant 29/03, 
and by the United States-Israel Binational Science 
Foundation (BSF) under grant 2002020.

\end{fmffile}
\end{document}